# A Taxonomy of Peer-to-Peer Based Complex Queries: a Grid perspective

Rajiv Ranjan, Aaron Harwood and Rajkumar Buyya
P2P Networks Group and GRIDS Laboratory
Department of Computer Science and Software Engineering
University of Melbourne, Victoria, Australia
{rranjan,aharwood,raj}@csse.unimelb.edu.au

October 31, 2018

**Abstract**

Grid superscheduling requires support for efficient and scalable discovery of resources. Resource discovery activities involve searching for the appropriate resource types that match the user's job requirements. To accomplish this goal, a resource discovery system that supports the desired look-up operation is mandatory. Various kinds of solutions to this problem have been suggested, including the centralised and hierarchical information server approach. However, both of these approaches have serious limitations in regards to *scalability, fault-tolerance and network congestion*. To overcome these limitations, organising resource information using Peer-to-Peer (P2P) network model has been proposed. Existing approaches advocate an extension to structured P2P protocols, to support the Grid resource information system (GRIS). In this paper, we identify issues related to the design of such an efficient, scalable, fault-tolerant, consistent and practical GRIS system using a P2P network model. We compile these issues into various taxonomies in sections 3 and 4. Further, we look into existing works that apply P2P based network protocols to GRIS. We think that this taxonomy and its mapping to relevant systems would be useful for academic and industry based researchers who are engaged in the design of scalable Grid systems.

## 1 Introduction

The last few years have seen the emergence of a new generation of distributed systems that scale over the Internet, operate under decentralized settings and are dynamic in their behavior (participants can leave or join the system). One such system is referred to as Grid Computing and other similar systems include P2P Computing, Semantic Web, Pervasive Computing and Mobile Computing. Grid Computing [51, 52] provides the basic infrastructure required for sharing diverse sets of resources including desktops, computational clusters, supercomputers, storage, data, sensors, applications and online scientific instruments. Grid Computing offers its vast computational power to solve grand challenge problems in science and engineering such as protein folding, high energy physics, financial modeling, earthquake simulation and climate/weather modeling, etc.

The notion of Grid Computing goes well beyond the traditional Parallel and Distributed Computing Systems (PDCS) as it involves various resources that belong to different administrative domains and are controlled by domain specific resource management policies. Furthermore, grids in general have evolved around complex business and service models where various small sites (resource owners) collaborate for computational and economic benefits. The task of resource management and application scheduling over a grid is complex due to resource heterogeneity, domain specific policies, dynamic environment, and various socio-economic and political factors.

Unlike traditional High-Performance Computing (HPC) systems such as mainframes or supercomputers whose control, usage and access were restricted to a single administrative domain, grid resources are topologically and administratively distributed over the Internet. There are different types of grid models [71] suitable for certain types of application service models including: (i) computational grids; (ii) data grids; and (iii) service grids. In this work, we mainly focus on the computational grids. Computational grids enable aggregation of different types of compute resources including clusters, supercomputers, desktops. In general, compute resources have two types of attributes: (i) static attributes such as the type of operating system installed, network bandwidth (both Local Area Network (LAN) and Wide Area Network (WAN) interconnection), processor speed and storage capacity (including physical and secondary memory); and (ii) dynamic attributes such as processor utilization, physical memory utilization, free secondary memory size, current usage price and network bandwidth utilization.

Grid computing assembles resources that are well managed, powerful and well connected to the Internet. Grids present a platform for grid participants (G's) to collaborate and coordinate resource management activities. Key GPs include the *producers* (grid resource-owners) and *consumers* (grid users). GPs have different goals, objectives, strategies, and supply and demand functions. GPs are topologically distributed and belong to different administrative domains. Controlled administration of grid resources gives an ability to provide a desired quality of service in terms of computational and storage efficiency, software or library upgrades. However, such controlled administration of resources gives rise to various social and political issues on which these resources are made available to the outside world. A resource owner invests a significant amount of money in establishing the resource, such as initial cost of buying, setting up, maintenance cost including hiring the administrator and expense of timely software and hardware upgrades. There is a complex diversity in terms of resources' usage policies, loads and availability. Resource owners in a grid behave as rational participants having distinct stake holdings with potentially conflicting and diverse utility functions. In this case, resource owners apply resource sharing policies that tend to maximize their utility functions [117, 48, 44, 83, 60].

## 1.1 The Superscheduling Process and Resource Indexing

The Grid superscheduling [101] problem is defined as: *"scheduling jobs across the grid resources such as computational clusters, parallel supercomputers, desktop machines that belong to different administrative domains"*. Superscheduling in computational grids is facilitated by specialized Grid schedulers/brokers such as the Grid Federation Agent [91], MyGrid [13], NASA-Superscheduler [102], Nimrod-G [12], GridBus-Broker [113], Condor-G [53] and workflow engines [118, 47]. Fig.1 shows an abstract model of a decentralised superscheduling system over a distributed query system. The superschedulers access the resource information by issuing lookup queries. The resource providers register the resource information through update queries. Superscheduling involves: (i) identifying and analyzing user's job requirements; (ii) querying GRIS [31, 66, 41, 119, 100, 15] for locating resources that match the job requirements; (iii) coordinating and negotiating Service Level Agreement (SLA) [86, 42, 39, 92]; and (iv) job scheduling. Grid resources are managed by their local resource management systems (LRMSes) such as Condor [74], Portable Batch System (PBS) [28], Sun Grid Engine (SGE) [57], Legion [36], Alchemi [75] and Load Sharing Facility LSF [8]. The LRMSes manage job queues, initiate and monitor their execution.

Traditionally, superschedulers including Nimrod-G, Condor-G and Tycoon [73] used services of centralized information services (such as R-GMA [121], Hawkeye [7], GMD [119], MDS-1 [49]) to index resource information. Under centralized organization, the superschedulers send resource queries to a centralized resource indexing service. Similarly, the resource providers update the resource status at periodic intervals using resource update messages. This approach has several design issues including: (i) highly prone to a single point of failure; (ii) lacks scalability; (iii) high network communication cost at links leading to the information server (i.e. network bottleneck, congestion); and (iv) the machine running the information services might lack the required computational power required to serve a large number of resource queries and updates.

To overcome the above shortcomings of centralized approaches, a hierarchical organization of information services has been proposed in systems such as MDS-3 [65] and Ganglia [4]. MDS-3 organizes Virtual Organization (VO) [51] specific information directories in a hierarchy. A VO includes a set of GPs that agree on common resource sharing policies. Every VO in grid designates a machine that hosts the information services. A similar approach has been followed in the Ganglia system, which is designed for monitoring resources status within a federation of clusters. Each cluster designates a node as a representative to the federated monitoring system. This node is responsible for reporting cluster status to the federation. However, this approach also has similar problems as the centralized approach such as one-point of failure, and does not scale well for a large number of users/providers.

## 1.2 Decentralized Resource Indexing

One of the well known examples of decentralized information services, or more precisely distributed search network, is the famous Web search engine Google [23]. The Google search network has deployed clusters of computers which are distributed worldwide. Each cluster has a few thousand processing nodes. A DNS-based load balancing scheme selects a cluster to process a user's search request depending on the user's geographic location. However, the servers which performs DNS based load-balancing are centralized component of the system. In case, these root name servers fail then the system may fail to function. In contrast, the current generation of P2P system does not rely on such central component hence are more scalable, available and fault-tolerant.

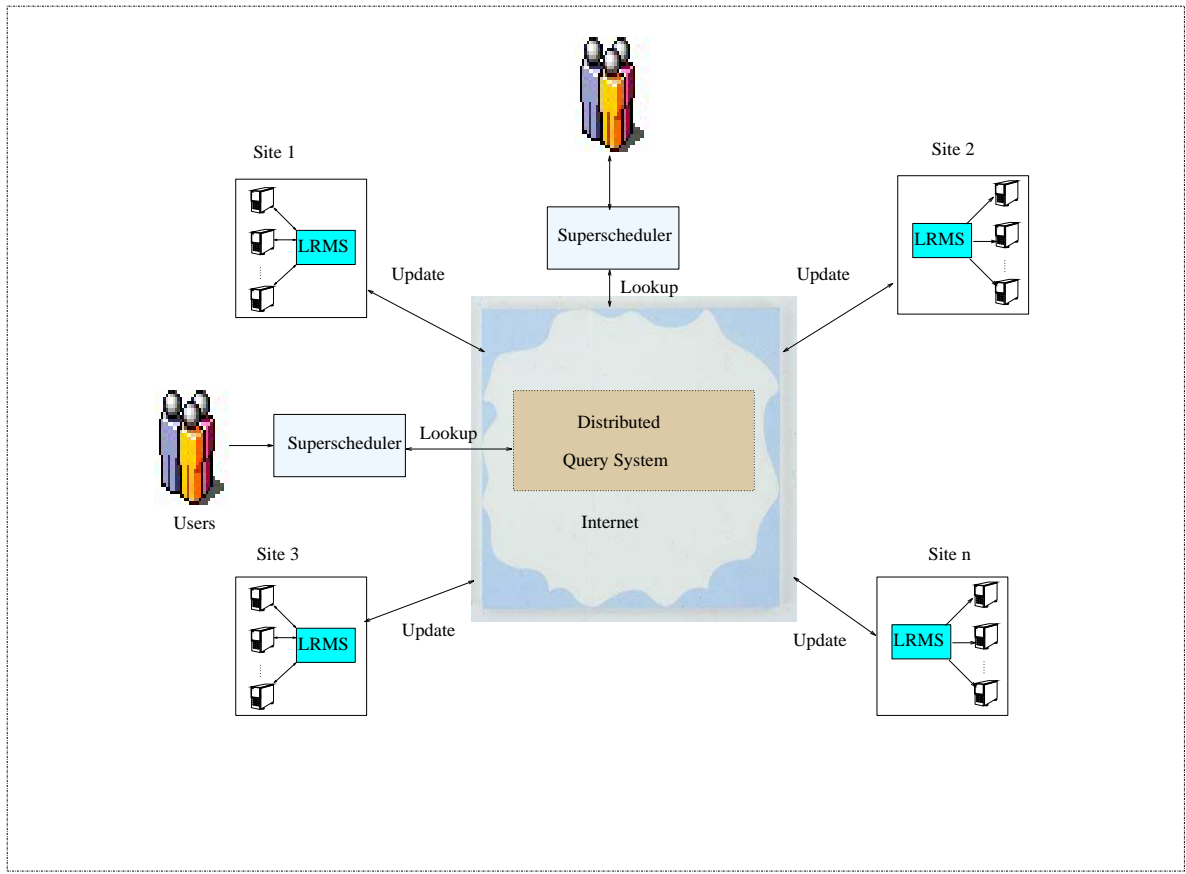

Figure 1: Superscheduling and resource queries

The load-balancing scheme focuses on minimizing round-trip time for the user's request, that also keeps into account the cluster's processing capability. Once the IP address of the relevant cluster is determined, the user's browser then sends a Hypertext Transport Protocol (HTTP) request to that cluster's web server (known as Google Web Server (GWS)). After that, all query processing activity is local to the cluster. The GWS machine coordinates query execution between index servers, document servers, Ad- servers and Spell-checker servers. When the query is resolved, the results are formatted as Hypertext Markup Language (HTML) and sent as a response message to the requesting browser. Fig. 2 shows the Google search network architecture.

However, a Google-like distributed search infrastructure can not be applied to Grid resource discovery. There are a number of reasons for that, including: (i) the information (documents) stored in the search engine databases are static, while the Grid resource attributes are dynamic in nature. Google web crawlers update the content for a web page every few weeks. While the attributes for a Grid resource may change in an interval of a few seconds. (ii) the semantics of two queries, i.e grid resource query and user's document search query are fundamentally different; (iii) resolving the grid resource queries is more complex and to an extent different than one applied for document indexing; and (iv) finally setting up a Google-like infrastructure for GRIS can be economically demanding. The business model for Google is based on the revenue earned from the online-advertising. Various service providers advertise their content description on web pages. Google earns money when the users visit those advertised web pages. However, such business model can not be applied for the web pages that may index grid resources, as the common search engine users do not directly access these pages.

Recently, proposals for decentralizing GRIS have gained significant momentum. The decentralization of GRIS

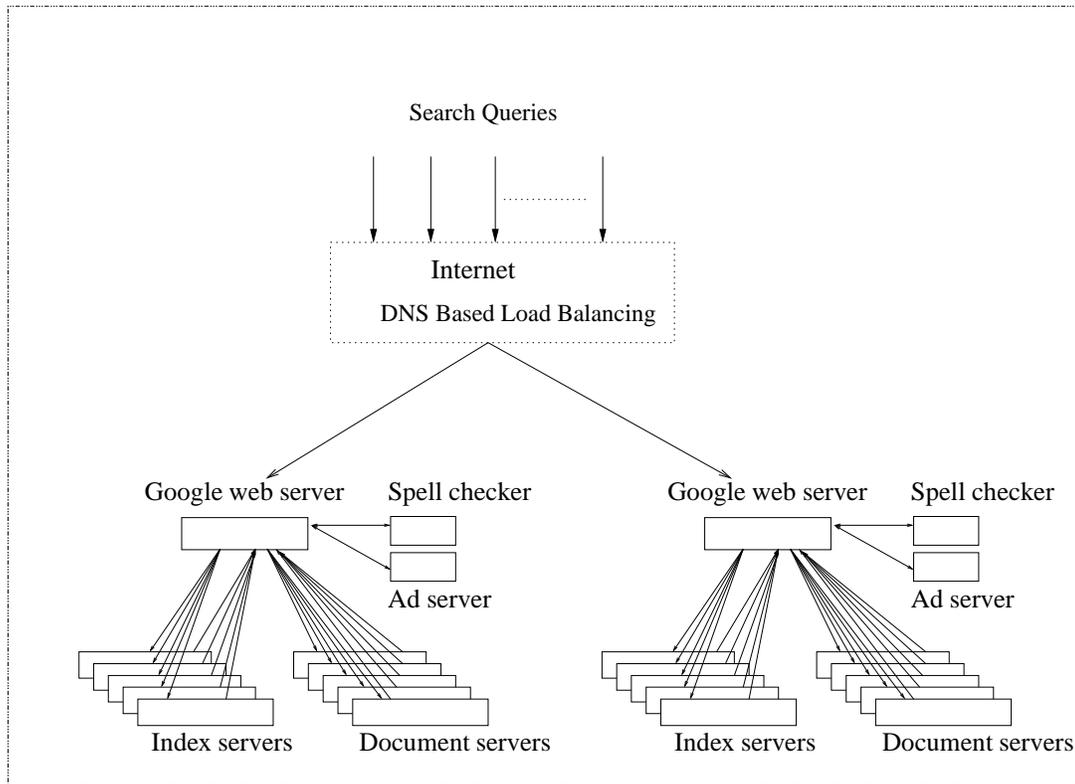

Figure 2: Google query-serving architecture

can solve grid resource fragmentation and other issues related to current centralized and hierarchical organizations. A distributed system configuration is considered as decentralized *"if none of the participants in the system is more important than others, in case one of the participant fails then it is neither more or less harmful to the system than caused by the failure of any other participant in the system"*. An early proposal for decentralizing Grid information services was made by Iamnitchi and Foster [66]. The work proposed a P2P based approach for organizing the MDS directories in a flat, dynamic P2P network. The work envisages that every VO maintains its information services and makes it available as part of a P2P based network. In other words, information services are the peers in a P2P network based coupling of VOs. Application schedulers in various VOs initiate a resource look-up query which is forwarded in the P2P network using flooding (an approach similar to one applied in the unstructured P2P network Gnutella [5]). However, this approach has a large volume of network messages generated due to flooding. To avoid this, a Time to Live (TTL) field is associated with every message, i.e. the peers stop forwarding a query message once the TTL expires. To an extent, this approach can limit the network message traffic, but the query search results may not be deterministic in all cases. Thus, the proposed approach can not guarantee to find the desired resource even though it exists in the network.

Recently, organizing a GRIS over structured P2P networks has been widely explored. Structured P2P networks offers deterministic query search results with logarithmic bounds on network message complexity. Structured P2P look-up systems including Chord [106], CAN [93], Pastry [96] and Tapestry [122] are primarily based on Distributed Hash Tables (DHTs). DHTs provide hash table like functionality at the Internet scale. It is widely accepted that they are the building blocks for next-generation large scale decentralized systems. Some of the example distributed systems include distributed databases [64], resource discovery systems [22, 37, 109, 100] and distributed storage systems [43]. A DHT is a data structure that associates a key with data. Entries in the hashtable are stored as a (key,data) pair. The data can be looked up if the corresponding key is known. DHTs have been proven to be scalable, fault-tolerant, self-organizing and robust. Current implementations of DHTs are known to be efficient for single-dimensional queries [64] such as "find all resources that match the given search point". In this case, distinct attribute values are specified for resource attributes. Extending DHTs to support multi-dimensional range queries such as finding all resources that overlap a given search space is a complex problem. Range queries are based on range of values for attributes rather than on a specific value. Current works including [37, 109, 100, 31, 40, 15, 84, 26, 89, 104] have studied and proposed different solutions to this problem.

## 1.3 Conceptual Design of a Distributed Resource Indexing System

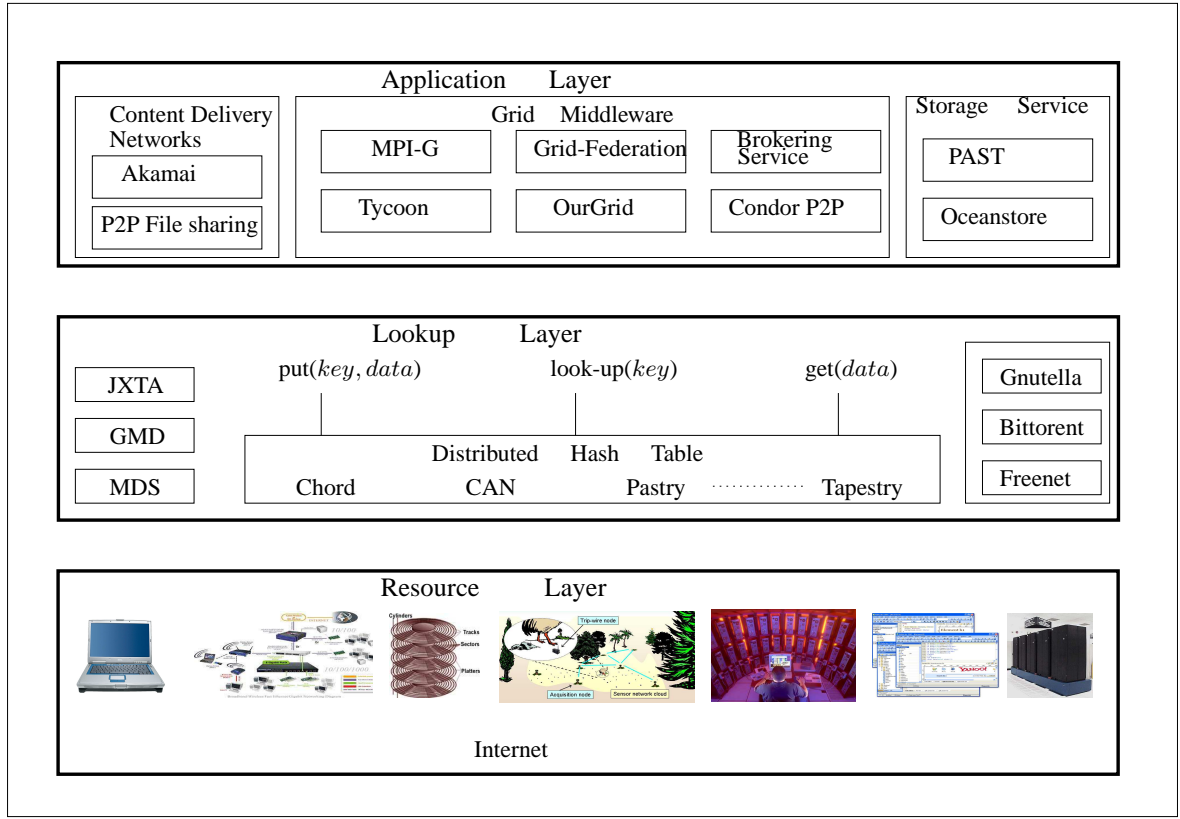

Figure 3: Distributed resource indexing: a layered approach

A layered architecture to build a distributed resource indexing system is shown in Fig. 3. The key components of the Internet-based resource indexing system includes:

- **Resource layer:** This layer consists of all globally distributed resources that are directly connected to the Internet. The range of resources includes desktop machines, files, supercomputers, computational clusters, storage devices, databases, scientific instruments and sensor networks. The computational resources can run variants of operating systems ( such as UNIX or Windows ) and queuing systems (such as Condor, Alchemi, SGE, PBS,LSF).

- **Lookup layer:** This layer offers core services for indexing resources at the Internet scale. The main components at this layer are the middlewares that support Internet-wide resource look-ups. Recent proposals at this layer have been utilizing structured P2P protocols such as Chord, CAN, Pastry and Tapestry. These are more commonly referred to as DHTs. DHTs offer deterministic search query performance while guaranteeing logarithmic bounds on the network message complexity. Other, middlewares at this layer includes JXTA [114], Grid Market Directory (GMD) [119] and unstructured P2P substrates such as Gnutella [5] and Freenet [3].

- **Application layer:** This layer includes the application services in various domains including: (i) Grid computing; (ii) distributed storage; (iii) P2P networks; and (iv) Content Delivery Networks (CDNs) [99], [87]. Grid systems including Condor-Flock P2P [30] uses services of Pastry DHT to index condor pools distributed over the Internet. Grid brokering system such as Nimrod-G utilizes directory services of Globus [50] for resource indexing and superscheduling. The OurGrid superscheduling framework incorporates JXTA for enabling communication between OGPeers in the network. Distributed storage systems including PAST [46] and OceanStore [72] utilizes services of DHTs such as Pastry and Tapestry for resource indexing.

## 1.4 Paper Organization

The rest of the paper is organized as follows. Section 2 summarizes the approaches that are based on a non-P2P resource organization model, specifically centralized and hierarchical network models. Section 3 presents taxonomies related to general computational resources' attributes, look-up queries and organization model. In section 4, we present taxonomies for P2P network organization, data distribution mechanism and query routing mechanism. Section 4.1 includes a brief summary of various structured P2P routing substrates that are utilized for organizing new generation GRIS systems. Section 5 summarizes various algorithms that model GRIS onto the P2P network model. Finally, we end this paper with discussion on open issues in section 6 and then conclude in section 7.

## 2 The State of Art in GRIS

The work in [120] presents a comprehensive taxonomy on existing centralized and hierarchically organized GRIS's. We summarize this work here and classify existing systems according to the proposed taxonomy in Table 1. The proposed taxonomy is based on the Grid Monitoring Architecture (GMA) [110] put forward by the Global Grid Forum (GGF). The main components of GMA are: (i) producer-daemon that monitors and publishes resource attributes to the registry; (ii) consumer-superschedulers that query the registry for resource information; (iii) registry-a service or a directory that allows publishing and indexing of resource information; (iv) republisher: any object that implements both producer and consumer functionality; and (v) schema repository-holds details such as type and schema about different kinds of events that are ambient in a GRIS. The work defines a scope-oriented taxonomy of existing GRIS. The systems are identified depending on the provision and characteristics of its producers and republishers.

Table 1: Summarizing centralized and hierarchical GRIS

| Level 0 | Level 1 | Level 2 | Level 3 |
|---|---|---|---|
| MapCenter [29], GridICE [14] | Autopilot [94] | CODE [103], GridRM [19], Hawkeye [7], HBM [105], Mercury [21], NetLogger [61], NWS [116], OCM-G [115], Remos [45], SCALEA-G [112] | Ganglia [4], Globus MDS [41], MonALISA [82], Paradyn [79], RGMA [121] |

- Level 0 (Self-Contained Systems): The resource consumers are directly informed of various resource attribute changes by the sensor daemon (a server program attached to the resource to monitor its attributes). The sensors are directly attached to the resource. The notification process may take place in an offline or an online setting. In the online case, the sensors locally store the resource metrics, which can be accessed in an application specific way. These systems normally offer a browsable web interface that provides interactive access to HTML-formatted information. These systems do not provide any kind of producer application programming interface (API), thus lacking any programming support that can enable automatic distribution of events to remotely located applications. Systems including MapCenter [29] and GridICE [14] belong to level 0 resource monitoring systems.

- Level 1 (Producer-Only Systems): Systems in this category have event sensors hosted on the same machine as the producer, or the sensor daemon functionality is provided by the producer itself. Additionally, these systems provide API interface at the resource level (producer level), hence they are easily and automatically accessible from remote applications. In this case, there is no need to browse through the web interface in-order to gather resource information. Systems including Autopilot [94] belong to the level 1 category of monitoring systems.

- Level 2 (Producers and Republishers): This category of system includes a republisher attached to each producer. The republisher of different functionality may be stacked upon each other but only in a predefined way. The only difference from Level 1 systems being the presence of a republisher in the system. Systems including GridRM [19], CODE [103] and Hawkeye [7] are level 2 systems.

- Level 3 (Hierarchies of Republishers): This category of system allows for the hierarchical organization of republishers in an arbitrary fashion which is not supported in Level 2 systems. In this arrangement, every node collects and processes events from its lower level producers and republishers . These systems provide better scalability than Level 0, Level 1 and Level 2 systems. Systems such as MDS-3 [41] belong to this category.

The taxonomy also proposes three other dimensions/qualifiers to characterize the existing systems. They include: (i) multiplicity-this qualifier refers to the scalability aspect (organization of the republisher in a Level 2 system) of a GRIS. A republisher can be completely centralized, or distributed with support of replication ; (ii) type of entities-types of resources indexed by the GRIS. These resource types include hosts, networks, applications and generic. A generic resource type at least supports event for hosts and network types; and (iii) stackable-denotes whether the concerned GRIS can work on top of another GRIS.

## 3 Resource Taxonomy

The taxonomy for a computational grid resource is divided into the following (refer to Fig. 4): (i) resource organisation; (ii) resource attribute; and (iii) resource query.

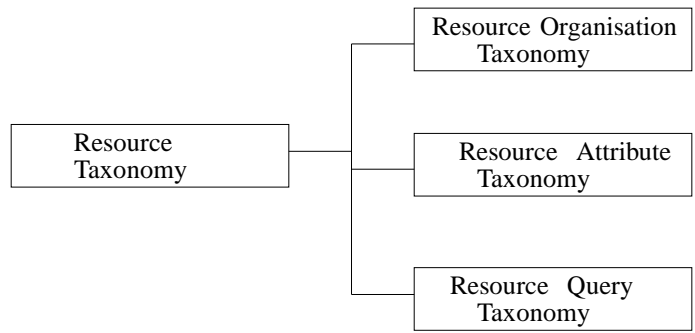

Figure 4: Resource taxonomy

### 3.1 Resource/GRIS organisation taxonomy

The taxonomy defines GRIS organization as (refer to Fig. 5) :

- Centralized: Centralization refers to the allocation of all query processing capability to single resource. The main characteristics of a centralized approach include control and efficiency. All look-up and update queries are sent to a single entity in the system. GRISes including RGMA [121] and GMD [119] are based on centralized organization.

- Hierarchical: A hierarchical approach links GRIS's either directly or indirectly, and either vertically or horizontally. The only direct links in a hierarchy are to a parent of a child. A hierarchy usually forms a tree like structure. GRIS system including MDS-3 [41] and Ganglia [4] are based on this network model.

- Decentralized: No centralized control, complete autonomy, authority and query processing capability is distributed over all resources in the system. The GRIS organized under this model is fault-tolerant, self-organizing and is scalable to large number of resources. More details on this organization can be found in section 4.

There are four fundamental challenges related to different organization models including: (i) scalability; (ii) adaptability; (iii) availability; and (iv) manageability. Centralized models are easy to manage but do not scale well. When network links leading to the central server get congested or fail, then the performance suffers. Hence, this approach may not adapt well to dynamic network conditions. Further, it presents a single point of failure, so overall availability of the system degrades considerably. Hierarchical organization overcomes some of these limitations including scalability, adaptability and availability. However, these advantages over a centralized model comes at the cost of overall system manageability. In this case, every site specific administrator has to periodically ensure the functionality of their local daemons. Further, the root node in the system may present a single point failure similar to the centralized model. Decentralized systems, including P2P, are coined as highly scalable, adaptable to network conditions and highly available. But manageability is a complex task in P2P networks as it incurs a lot of network traffic.

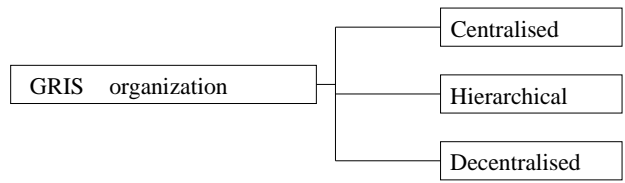

Figure 5: Resource organization taxonomy

## 3.2 Resource Attribute Taxonomy

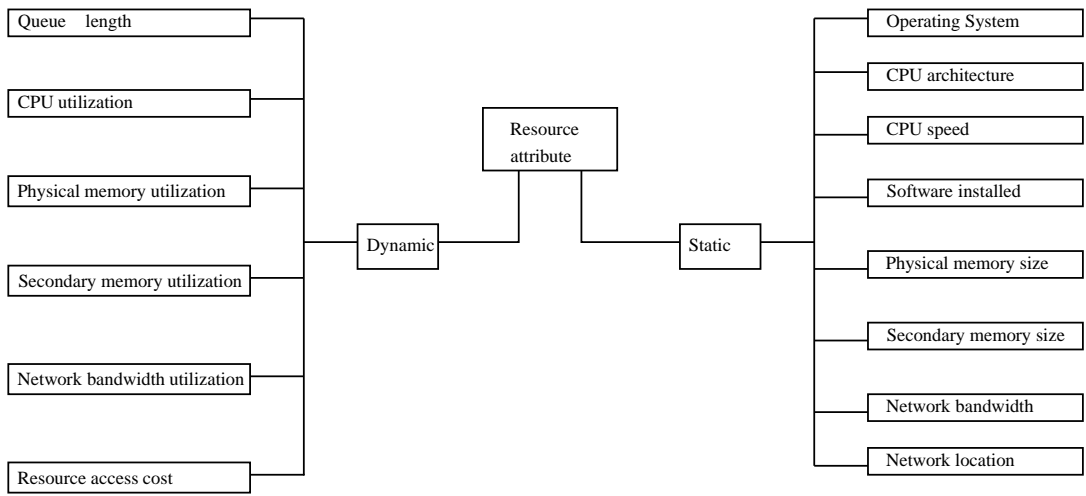

Figure 6: Resource attribute taxonomy

A compute grid resource is described by a set of attributes which is globally known to the application superschedulers. The superscheduler which is interested in finding a resource to execute a user's job issues queries to GRIS. The queries are a combination of desired attribute values or their ranges, depending on the user's job composition. In general, compute resources have two types of attributes: (i) static or fixed value attributes such as: type of operating system installed, network bandwidth (both LAN and WAN interconnection), network location, CPU speed, CPU architecture, software library installed and storage capacity (including physical and secondary memory); and (ii) dynamic or range valued attributes such as CPU utilisation, physical memory utilisation, free secondary memory size, current usage price and network bandwidth utilisation. Figure 6 depicts the resource attribute taxonomy.

### 3.3 Resource Query Taxonomy

The ability of superschedulers such as MyGrid, Grid-Federation Agent, , NASA-Superscheduler, Nimrod-G, Condor-Flock P2P to make effective application scheduling decision is directly governed by efficiency of GRIS. Superschedulers need to query a GRIS to compile information about resource's utilisation, load and current access price for formulating the efficient schedules. Further, a superscheduler can also query a GRIS for resources based on selected attributes such as nodes with large amounts of physical and secondary memory, inter-resource attributes such as network latency, number of routing hops or physical attributes such as geographic location. Similarly, the resource owners query GRIS to determine supply and demand pattern and accordingly set the price. The actual semantics of the resource query depends on the underlying Grid superscheduling model or Grid system model. Various grid system models [71] include computational grids, data grids and service grids.

#### 3.3.1 Resource Query Type

Superscheduling systems require two basic types of queries: (i) resource look-up query (RLQ) ; and (ii) resource update query (RUQ). An RLQ is issued by the superscheduler to locate resources matching the user's job requirements, while an RUQ is an update message sent to GRIS by the resource owner about the underlying resource conditions. In Condor-flock P2P system, flocking requires sending RLQs to remote pools for resource status and the willingness to accept remote jobs. Willingness to accept remote jobs is a policy specific issue. After receiving the RLQ message, the contacted pool manager replies with RUQ that includes the job queue length, average pool utilization and number of resources available. The distributed flocking is based on the P2P query mechanism. Once the job is migrated to the remote pool, basic matchmaking [90] mechanism is applied for resource allocation. In Table 2, we present RLQ and RUQ queries in some well-known superscheduling systems.

#### 3.3.2 An Example Superscheduling Resource Query

In this section we briefly analyze the superscheduling query composition in a superscheduling system called Tycoon [73]. Tycoon applies market-based principles, in particular the auction mechanism, for resource management. The auctions are completely independent without any centralised control. Every resource owner in the system coordinates its own auction for local resources. The Tycoon system provides a centralised Service Location Service (SLS) for superschedulers to index resource auctioneers' information. Auctioneers register their status with the SLS every 30 seconds. If an auctioneer fails to update its information within 120 seconds then the SLS deletes its entry. Application level superschedulers contact the SLS to gather information about various auctioneers in the system. Once this information is available, the superschedulers (on behalf of users) issue bids for different resources (controlled by different auctions), constrained by resource requirement and available budget. A resource bid is defined by the tuple $(h, r, b, t)$ where $h$ is the host to bid on, $r$ is the resource type, $b$ is the number of credits to bid, and $t$ is the time interval over which to bid. Auctioneers determine the outcome by using a bid-based proportional resource sharing economy model.

Auctioneers in the Tycoon superscheduling system send an RUQ to the centralised GRIS (referred to as service local services). The update message consists of the total number of bids currently active for each resource type and the total amount of each resource type available (such as CPU speed, memory size, disk space). An auctioneers RUQ has the following semantics with example values:

$$total\ bids = 10\ \&\&\ CPU\ Arch = \text{``pentium''}\ \&\&\ CPU\ Speed = 2\ GHz\ \&\&\ Memory = 512$$

Similarly, the superscheduler, on behalf of the Tycoon users, issues an RLQ to the GRIS to acquire information about active resource auctioneers in the system. A user resource look-up query has the following semantics:

$$return\ auctioneers\ whose\ CPU\ Arch = \text{``i686''}\ \&\&\ CPU\ Speed \geq 1\ GHz\ \&\&\ Memory \geq 256$$

Table 2: Resource query in superscheduling systems

| System Name | Resource Lookup Query | Resource Update Query | GRIS Model |
|---|---|---|---|
| Condor-Flock P2P | Query remote pools in the routing table for resource status and resource sharing policy | Queue length, average pool utilization and number of resources available | Decentralised |
| Grid-Federation | Query decentralised federation directory for resources that matches user's job QoS requirement (CPU architecture, no. of processors, available memory, CPU speed) | Update resource access price and resource conditions (CPU utilisation, memory, disk space, no. of free processors) | Decentralised |
| Nimrod-G | Query GMD or MDS for resources that matches jobs resource and QoS requirement | Update resource service price and resource type available | Centralised |
| Condor-G | Query for available resource using Grid Resource Information Protocol (GRIP), then individual resources are queried for current status depending on superscheduling method | Update resource information to MDS using GRRP | Centralised |
| Our-Grid | MyPeer queries OGPeer for resources that match user's job requirements | Update network of favors credit for OurGrid sites in the community | Decentralised |
| Gridbus Broker | Query GMD or MDS for resources that matches jobs resource and QoS requirement | Update resource service price and resource type available | Centralised |
| Tycoon | Query for auctioneers that are currently accepting bids and matches user's resource requirement | Update number of bids currently active and current resource availability condition | Centralised |
| Bellagio | Query for resources based on CPU load, available memory, inter-node latency, physical and logical proximity | Update resource conditions including CPU, memory and network usage status | Decentralised |
| Mosix-Grid | Information available at each node through *gossiping algorithm* | Update CPU usage, current load, memory status and network status | Hierarchical |

In Fig. 7, we present the taxonomy for GRIS RLQ and RUQ. In general, the queries [95] can be abstracted as lookups for objects based on a single dimension or multiple dimensions. Since, a grid resource is identified by more than one attribute, the RLQ or RUQ are always multidimensional. Further, both the single dimension and multidimensional query can specify different kinds of constraints on the attribute values. If the query specifies a fixed value for each attribute then it is referred to as a *multi-dimensional point query*. However, in case the query specifies a range of values for attributes, then it is referred to as a *multi-dimensional window query*. Depending on how values are constrained and searched for, these queries are classified as:

- Exact match query: The query specifies the desired values of all resource attributes sought. For example, Architecture='x86' and CPU-Speed='3 Ghz' and type='SMP' and price='2 Grid dollars per second' and RAM='256 MB' and No. of processors=10 and Secondary free space='100 MB' and Interconnect bandwidth='1 GB/s' and OS='linux'. (Multiple Dimension Exact Match Query).

- Partial match query: Only selected attribute values are specified. For example, Architecture='sparc' and type='SMP' and No. of processors=10. (Multiple Dimension Partial Match Query).

- Range queries: Range values for all or some attributes are specified. For example, Architecture='Macintosh' and

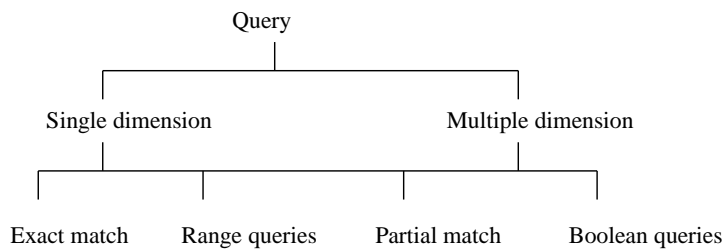

Figure 7: Resource query taxonomy

type='Cluster' and '1 GHz' $\leq$ CPU-Speed $\leq$ '3 GHz' and '512MB' $\leq$ RAM $\leq$ '1 GB'. (Multiple Dimension Range Query).

- Boolean queries: All or some attribute values satisfying certain boolean conditions. Such as, ((not RAM $\leq$ '256 MB') and not No. of processors $\leq$ 5). (Multiple Dimension Boolean Query).

## 4 P2P Taxonomy

Before presenting the taxonomy, we briefly describe some of the existing structured P2P systems in the following section. These systems have been extensively used in the design and development of P2P based GRIS.

### 4.1 Structured P2P Systems: Summary

P2P resource sharing paradigm started with famous music file sharing systems such as Gnutella [5], Freenet [3] and Napster [9]. These unstructured P2P systems were based on inefficient message routing and object look-up scheme. Napster system applied a central index server that maintained location information of data items currently available in the network. Every interested peer are required to query the central index server to obtain the location of server that maintains the desired content. Although this approach has look-up complexity of $O(1)$, it is seriously prone to one point failure and performance bottleneck (query overload). Another message routing scheme inherently applied in Gnutella system requires interested peers to broadcast the query message to all of its neighbors. The recipient peer indexes its local database. In case the desired information is found, the peer directly responds with the item. Otherwise, it forwards the query message its own neighbors. A Time-to-Live (TTL) parameter is associated with each query to avoid request loops. However, this query flooding scheme does not scale well, as significant overhead involved in the network bandwidth and compute CPU cycle consumption of intermediate peers. To negate the flooding approach, recent systems such as Kazza [59], Grokster [6] adopted hierarchical arrangement of nodes in the overlay network. In this scheme, every search start at the root node that traverses a single search path down to the peer that contains the desired data. Although this approach is more efficient than flooding, still it suffers from bottleneck and one point failure.

Structured P2P systems have recently emerged as the next generation infrastructure to build large scale distributed systems. They offer enhanced efficiency, scalability and fault-tolerance compared to their progeny's such as Gnutella, Napster, Freenet and Kazza. They include Chord [106], Content Addressable Network (CAN) [93], Pastry [96] and Tapestry [122] etc. In general, DHTs implement one core functionality: given that the key $k$ is routed to an appropriate node $n$ in the overlay network that holds the given key. DHTs can resolve lookups for a 1-dimensional object within a logarithmic overlay routing hops.

Inherent in the design of a DHT are the following issues [20]: (i) generation of node-ids and object-ids, called keys, using cryptographic hash functions such as SHA-1 [11]. The objects and nodes are mapped on the overlay network depending on their key value. Each node is assigned responsibility for managing a small number of objects; (ii) building up routing information (routing tables) at various nodes in the network. Each node maintains the network location information of a few other nodes in the network; and (iii) an efficient look-up query resolution scheme. Whenever a node in the overlay receives a look-up request, it must be able to resolve it within acceptable bounds such as in $\log n$ time. This is achieved by routing the look-up request to the nodes in the network that are most probable to store the information about the desired object. Such probable nodes are identified by using the routing table entries.

Though at the core DHTs are similar, still there exists substantial differences in the actual implementation of algorithms including the overlay network construction (network graph structure), routing table maintenance and node join/leave handling. The performance metrics for evaluating a DHT include fault-tolerance, load-balancing, efficiency of lookups and inserts and proximity awareness. We present a detailed discussion about the DHTs with focus on these core issues.

- CAN

    The core design of CAN is based on a virtual $d$-dimensional Cartesian coordinate space or $d$-torus. The coordinate space is completely logical in abstraction having no resemblance to any physical coordinate system. The entire coordinate space is divided into various hyper-rectangles called zones. Every node in the CAN overlay network owns a particular zone within the overall space and is identified by a zone's coordinate value. Fig.8. shows a 2-dimensional $[0,1] \times [0,1]$ coordinate space divided among 5 nodes. Each node maintains a routing table that contains the entry (IP address) for all its neighbors. Two nodes are considered to be neighbors if their coordinate spans overlap along $d-1$ dimensions and meet along one dimension. The virtual coordinate space stores the $(key, value)$ pairs as follows: a key (1-dimensional object) is deterministically mapped to a point in the space using the uniform hashing function. The node under whose zone the point maps is held responsible for that corresponding $(key, value)$ pair. To locate a key, a CAN node routes the query message along the path that leads to the key's destination. This path is an approximate straight line in the coordinate space, from the node that initiates the query to the node storing the key. The message routing/forwarding mechanism simply follows a greedy forwarding to the neighbor with coordinates closest to the destination coordinates. Fig.8. shows a sample routing path from node $E$ to the node $A$. In a $d$ dimensional space partitioned into $n$ equal zones, the average routing path complexity is $O(d\ n^{1/d})$ hops and individual nodes maintain $O(2\ d)$ neighbors network addresses. Note that, when $d = \log n$, CAN has $O(\log n)$ neighbor and the total routing cost is $O(\log n)$. Note that, all our logarithmic notations are to the base 2, unless explicitly specified.

    To join the CAN network, a node $k$ randomly chooses a point $p = (x, y)$ in the coordinate space and sends a join request message destined for point $p$. Note that, node $k$ discovers the existing CAN network through some predefined bootstrap node. A bootstrap node maintains a partial list of CAN nodes currently ambient in the network. The message is forwarded in the network using the standard CAN routing scheme until the node $m$ is reached in whose zone $p$ lies. Following this, the node $m$ splits its zone into two equal parts and relinquishes the ownership of one of them to the new node $k$. The new node builds its routing table by simply copying the subset of entries from $m$'s table and one additional entry for node $m$ itself. The new node $k$ sends update message to all of its neighbor. This facilitates the neighbors to update their routing table with the new node. When nodes leave a CAN, they handover their zone to one their neighbors. In case, the merging of two zones creates a valid new zone, the two zones are combined into a larger zone. If not, the neighbor node will temporarily handle both the zones. If any node fails, then CAN implements the zone assignment protocol that relinquishes the control of a new zone to the neighbor with smallest zone.

    CAN algorithm proposes various design techniques to reduce the overall routing latency and routing fault-tolerance. First design improvement advocates increasing the number of dimensions of the coordinate space. In this case, every node has more entries in the routing table (i.e. more number of neighbors). This improves the routing fault-tolerance as a node now has more routing options. The second design observation suggests maintaining multiple, independent coordinate spaces with each node owning different zones in different spaces. This is refereed as having multiple realities. A node in a CAN of $r$ realities is assigned $r$ zones and on every reality it holds independent routing table or neighbor sets. This replication scheme applies to the data as well. So this scheme not only improves the routing but also the data availability. Finally, CAN improves the look-up latency by having each node measuring the network round-trip time (RTT) to each of its neighbors. When performing a look-up operation, the message is routed to the neighbor with maximum ratio of progress towards the probable destination node to the RTT. This scheme always opts for low latency paths in the network, and avoids unnecessary long hops.

- Pastry

    The Pastry overlay protocol organises the nodes into a 1-dimensional circular-ID space. Each node is assigned a unique 128-bit node identifier and the node Ids are generated through a SHA-1 hash of the node's IP address or public key. The node Ids are uniformly distributed in 128-bit node Id space and conceptually it indicates the

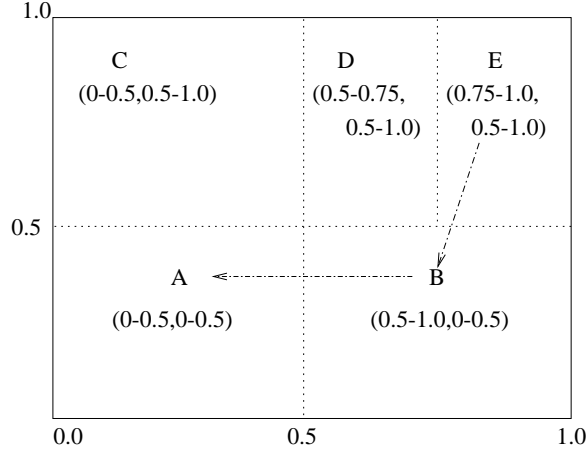

Figure 8: 2-dimensional CAN space with 5 nodes

corresponding node's position on the overlay circle. Each node in the overlay maintains the following routing structure: (i) a routing table $R$; (ii) a neighborhood set $M$; and (iii) a leaf set $L$.

The routing table $R$ has $\log_{2^b} N$ rows and $2^b - 1$ columns. Set of entries $(nodeID, IPaddress)$ in the $i$th row corresponds to the nodes that share same prefix in first $i$ digits the with present node's id. While the $i+1$th digit has one of the $2^b - 1$ (where $b$ is a configuration parameter) possible values other than $i+1$th digit in the present node's id. The leaf set $L$ maintains the set of nodes whose ids are numerically closer to the present node's id. $L$ is portioned into two subsets, with first subset contain the nodes whose ids are larger while the second subset constitute the nodes whose ids are smaller than the present node's id. The neighborhood set $M$ includes the entries for nodes that are topologically near in the Internet. Typically the size of $L$ and $M$ are $2^b$ or $2 \times 2^b$.

To route a message with a given key $k$, a node first matches $k$ against the entries in $L$. If $k \in L$, then the message is directly routed to the node whose id is closest to $k$. However, if the message key is not covered by the leaf set, then the routing table entries are used to route the message to an intermediate node that shares the common prefix with the key at-least one more digit than the present node's id or at-least is more numerically closer. The overlay routing scheme guarantees average message routing complexity of $\log_{2^b} N$, subject to accurate routing table state and no recent node failure or departures. Fig. 9 shows the example message routing from node with id 45600000 to node with id 23680000.

The node join operation involves initiating contact with one of the nodes already in the Pastry overlay. Typically, this node is referred as the bootstrap node and can be located using "expanding ring" IP multicast. The new node $X$ sends sends the overlay $join$ message (SHA-1 hash of IP address or public key) to bootstrap node $A$ destined for some node $Z$ that is numerically closest to the new node's id. Node $A$ routes the $join$ message using the standard pastry routing scheme. Every intermediate nodes on the routing path sends its state table ($R$, $L$ $and$ $M$) to the newly joined node $X$. The node $X$ inspects these state tables, if required requests additional messages and initializes its local state tables. This process is repeated for each node encountered along the path from node $A$ to node $Z$. Once node $X$ tables are ready, it broadcasts its local states to all the nodes in its $R$, $L$ $and$ $M$ set. Following this, every node in turn update their own state using this new information. Total message complexity involved in a node join operation is $O(\log_{2^b} N)$.

Pastry algorithm applies different techniques to counter the failed node in $R$, $L$ $and$ $M$ sets. In case a node $\in L$ set fails, then the node that owns $L$ contacts a live node in the overlay having the largest index on the side of the failed node. Once such node is discovered, it is asked to reply with its leaf set $L^n$ set. On arrival of this new $L^n$ set, the concerned node chooses an appropriate node from $L^n$ to repair its local $L$. To repair the routing table entry $R_l^d$, a new overlay node $R_l^i$ is contacted, such that $i \neq d$ in the same row and asks for that node's entry for $R_l^d$. However, if the node $R_l^i$ does not return a valid node then the node $R_{l+1}^d$, $i \neq d$ is contacted, thereby casting a wider net. Pastry also implements number of heuristics to facilitate proximity aware message routing. To do this, each node analyzes nodes in its $M$ set and re-initializes the routing table entries to reflect that each entry refers to a topologically *nearby* node with appropriate prefix, among all active nodes.

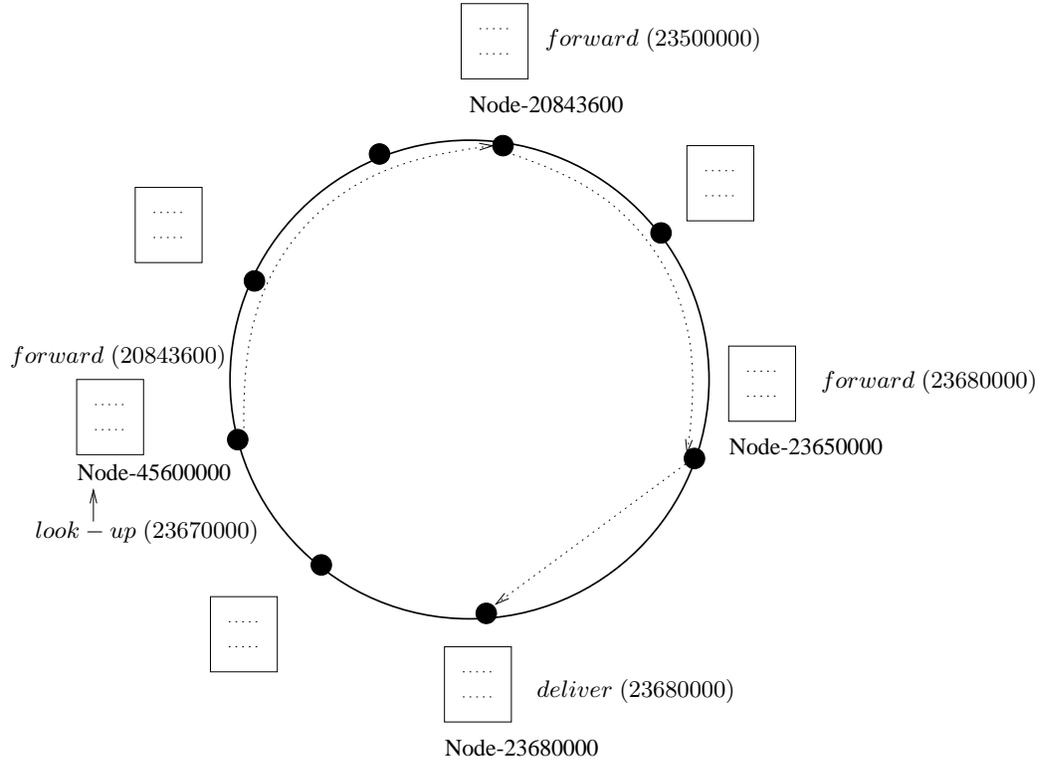

Figure 9: Example Pastry ring routing node id 45600000 to node id 2368000 (with 16-bit node id expressed as eight 2-bit values)

- Chord
  The Chord protocol organises the nodes and the data items into a 1-dimensional circular space. A consistent hashing such as SHA-1 is used to generate the corresponding identifier and key that form the basis for location of nodes and data items on the circular overlay. A node identifier is generated by hashing its IP address, while the data item's key is produced by hashing its unique name. Consistent hashing maps keys to nodes as follows: first nodes are ordered on an identifier circle of modulo $2^m$. Then key $k$ is assigned to that node which directly follows it in the identifier space. This node is called the successor node of $k$, denoted by $successor(k)$. Fig. 10 shows a Chord overlay network of size $2^6$. Chord controls the uniform distribution of keys to nodes by means of *virtual nodes*. In this case, every physical node has multiple virtual images with different Ids in the same Chord system. This ensures the uniform distribution of keys over more nodes in the system.

Each node in the overlay maintains information about $\log(n)$ other nodes in the network. This information is stored in a routing table more commonly referred as a *finger table* (FT). A node's finger table contains the IP address and corresponding identifier for each node. The index $i$ in a FT is computed using the formula $(n + 2^{i-1})$, where $n$ is the node identifier. These node ids are exponentially distributed on the space. Hence, in this case a node indexes more nodes that are near to it on the overlay than those far away. The $i$-th entry in the finger table corresponds to a node that succeeds the given node $n$ by at-least $2^{i-1}$ on the identifier circle. This power-of-two finger table entries ensures that the node can always forward the query to at-least half of the remaining ID space. As a result Chord lookups on average takes around $O(\log N)$ routing hops.

The query to look-up a key $k$ at a node $n$ is resolved iteratively. If the node $n$ can find the FT entry for the node $successor(k)$, then the query is resolved in $O(1)$ time complexity. However, in case FT does not directly indexes the $successor(k)$ then $n$ forwards the query to node $j$ whose ID is more closer than its own to $k$. The node $j$ in turn returns the node ID which is further closer to $k$. Thus by repeating this process (on the average $O(\log N)$), a node $n$ discovers the nodes with IDs closer and closer to $k$. Introduction of a new node $n$ to the network is facilitated through a node $l$ already part of the overlay. In this case, the node $l$ helps node $m$ to figure out its position in the identifier space along with its predecessor and successor nodes. The new node join process requires that few existing keys exchange hands (some of the keys owned by successor moves to the new node $n$). Further new node $n$ needs to be entered into the FTs of some existing nodes (in particular those nodes which indexes the key owned by the new node $n$).

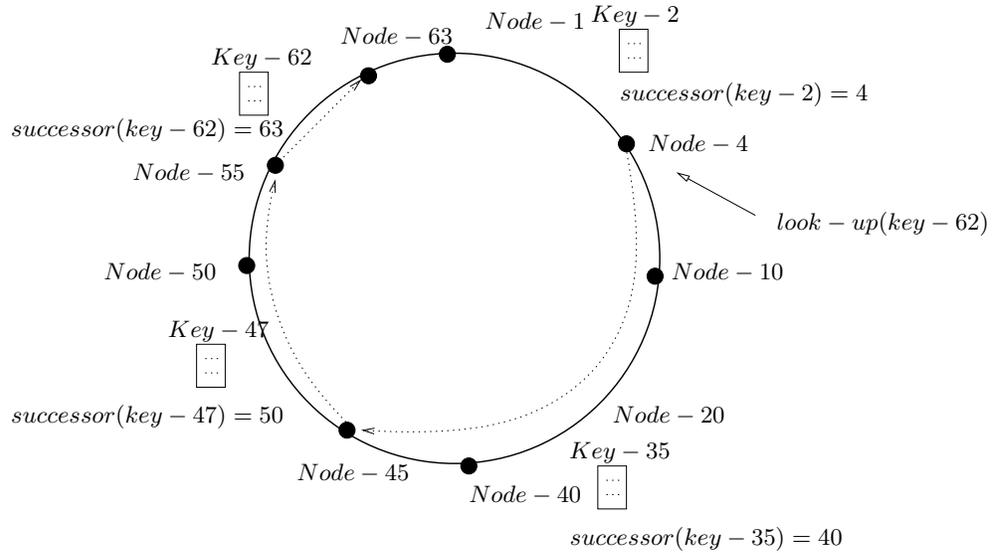

Figure 10: The routing path taken to resolve the look-up for key 62, initiated at node 4, in an circular ID space of size $2^6$. The arrows represent the path taken by the message from node 4 to node 64

In case, a node $q$ leaves an existing Chord overlay the keys owned by it needs to be reassigned to its present successor node. Accordingly, FTs entries of existing nodes need to be updated to reflect the change in key ownership. On the average $O(\log^2 N)$ message needs to be exchanged between nodes to fully accomplices node join or leave operation. In addition to this, Chord also runs a network stabilization procedure that keeps track of existing nodes' successor pointers up to date. Note that, maintaining correct successor pointer to a extent guarantees that no look-up operation will fail.

- Tapestry

The main design motivation of the Tapestry algorithm is the Plaxton Mesh (PM) [88]. The PM allows physical network proximity-aware object location and message routing in a arbitrarily-sized network. Each node in the PM maintains a small amount of routing information. Further, PM guarantees message delivery in near optimal time, from any point in the network. Every object and node is assigned a location independent name. Names are based on random fixed-length bit-sequences in a common base such as 40 hexadecimal digits representing 160 bits. The system guarantees an even distribution of objects and nodes over the space by applying a of hashing algorithm, such as SHA-1. The routing information at each node in the overlay is maintained in a table called the *neighbor map* (NM). The NM has multiple levels, where each level represents a matching suffix up to a digit position of the present node ID. The number of entries at each level is equal to the base of the ID space in the overlay. The $i$-th entry in the $j$-th level is the ID and location of the closest node which ends in "$i$"+ $suffix\ (n, j - 1)$. So, for example, the 9-th entry of the 4-th level for node $846AE$ is the node topologically nearest to it and whose ID ends with $96AE$. In case a message arrives from a node in the overlay that shares a suffix of at-least length $n$ with the final destination nodes, the next node for forwarding the given message is chosen which exist at level $n+1$ in the NM. Under ideal conditions, such as consistent NMs, the routing scheme guarantees successful delivery of messages within $O(\log_b(n))$ logical hops. To summarize, every destination node is the *root node* of its own tree, which is a unique spanning tree across all the nodes. Every leaf node can traverse a number of intermediate nodes en route to the root node. Fig. 11. shows an example Tapestry/Plaxton

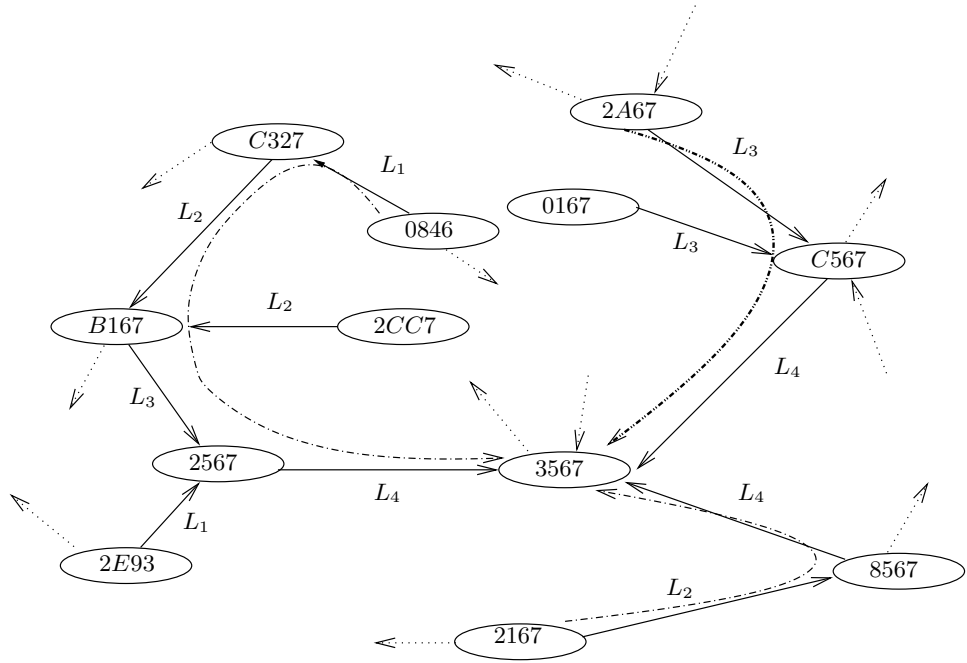

Figure 11: Example Plaxton/Tapestry routing: Sample routing path taken by message from node IDs 0846, 2A67 and 2167 to node ID 3567 in Plaxton mesh using hexadecimal digits having length 4.

mesh routing.

| P2P system | Overlay Structure | Network parameter | Routing table size | Routing complexity | join/leave overhead |
|---|---|---|---|---|---|
| Chord | Uni-dimensional, circular-ID space | $n$= number of nodes in the network | $\log(n)$ | $\log(n)$ | $(\log(n))^2$ |
| Pastry | Plaxton-style mesh structure | $n$= number of nodes in the network, $b$=base of the identifier | $\log_b(n)$ | $b \log_b(n) + b$ | $\log(n)$ |
| CAN | Multidimensional ID space | $n$= number of nodes in the network, $d$=number of dimensions | $2d$ | $d n^{1/d}$ | $2d$ |
| Tapestry | Plaxton-style mesh structure | $n$= number of nodes in the network, $b$=base of the identifier | $\log_b(n)$ | $b \log_b(n) + b$ | $\log(n)$ |

Table 3: Summary of the complexity of structured P2P algorithms

A node in the overlay (may be referred to as server node) publishes that it has an object $'O'$ by routing a message to the *"root node"* of $O$. The root node is a unique node in the overlay that is responsible for maintaining the root of the embedded unique spanning tree for object $O$. An object root or *surrogate* node is the node which matches the objects ID($I$) *in the greatest number of trailing bit positions*. Along the publish message routing path, each intermediate routing node maintains object's $O$ indexing information as the pair $(Object - ID(O), Server -$ $IDf(S)$). This index is a pointer to the actual object $O$ stored on server $S$. The look-up request for object $O$ is routed towards its root node. However, if the intermediate node on the look-up path happens to store the information object's index then the query is directly resolved with the ID of the concerned server returned to the requester. However, the root node for a object $O$ may act as one point failure and make the object invisible in the overlay.

In addition to basic PM's object routing and routing substrate, Tapestry implements various fault-tolerant procedure to counter server outages, link failures, and neighbor table corruption at any node. Link and server failures are detected using the TCP timeouts. Further, each Tapestry node maintains back-pointers to the nodes for which it is a neighbor. These back-pointers are utilized to send periodic heartbeat UDP messages ("Hello message"). By analyzing the frequency of such incoming messages, a overlay node can easily figure out the failed or corrupted nodes in its NM. Tapestry overcomes the one point failure of the root node by assigning multiple roots to each object $O$. This is accomplished by concatenating a small globally constant sequence of "salt" values such as 1, 2, 3 to each object ID, then result is hashed to link it to the appropriated roots. Note that, each of the salt values defines an independent *routing plane* for Plaxton-style location and routing. The look-up requests performs the same hashing process with the target object ID that generates same root node set.

Finally, Tapestry algorithm clearly distinguishes itself from other DHTs by focusing on network proximity routing. The routing scheme forwards query to those nodes with minimum network latency. In case, there exists multiple object copies, an effort is made to route to closest available replica. Experimental evaluation proves that Tapestry scheme indeed finds an approximate nearest copy of any sought Object. However, the proximity aware routing comes at the cost of increased complexity in routing data structures.

The taxonomy for P2P based GRIS is divided into the following (refer to Fig. 12): (i) P2P network organisation; (ii) data organisation; and (iii) query routing organisation.

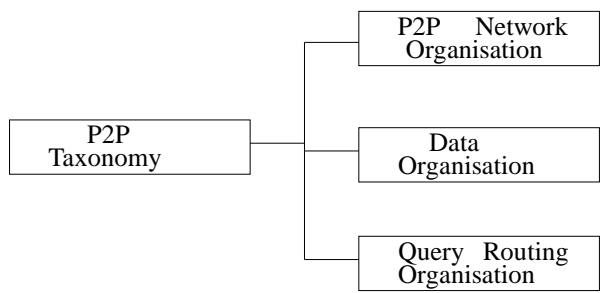

Figure 12: Peer-to-Peer network taxonomy

## 4.2 P2P Network Organisation

Fig. 13 shows the network organisation taxonomy of general P2P systems. The network organisation refers to how peers are logically structured from the topological perspective. Two categories are proposed in the P2P literature [80]: unstructured and structured. An unstructured system is based on a power law random graph model [25, 38]. This system does not put any constraint over placement of data items on peers and how peers maintain their network connections. Detailed evaluation and analysis of network models [69], [32] for the unstructured systems can be found in [77]. Unstructured systems including Napster, Gnutella and Kazza offer differing degrees of decentralization. The degree of decentralization means to what extent the peers can function independently with respect to efficient look-up and message routing. These degrees are categorised into: fully decentralised, partially decentralised and hybrid decentralised.

In a hybrid decentralised organisation, every peer depends on the centralised indexing server to obtain the index of the peer responsible for a particular data item. Example P2P systems in this category include Napster and BitTorrent [2]. Systems such as Kazza and FastTrack are partially decentralised in their organisation. Peers in these overlay networks form a structured overlay of super-peer architectures. Super-peers represent a class of peers which are relatively more important to the functioning of the overall system. Super-peers maintain state information of other super-peers in the network. In general, super-peers are nodes with high computational, storage and network communication ability. Simple peers send the look-up queries and indexes of data they store to the super-peer. In a fully decentralised organisation, all peers have equal functionality and there is no dependency on any specialised server such as super-peer or centralised indexing directory. Systems including Gnutella and Freenet belong to this category of unstructured P2P systems.

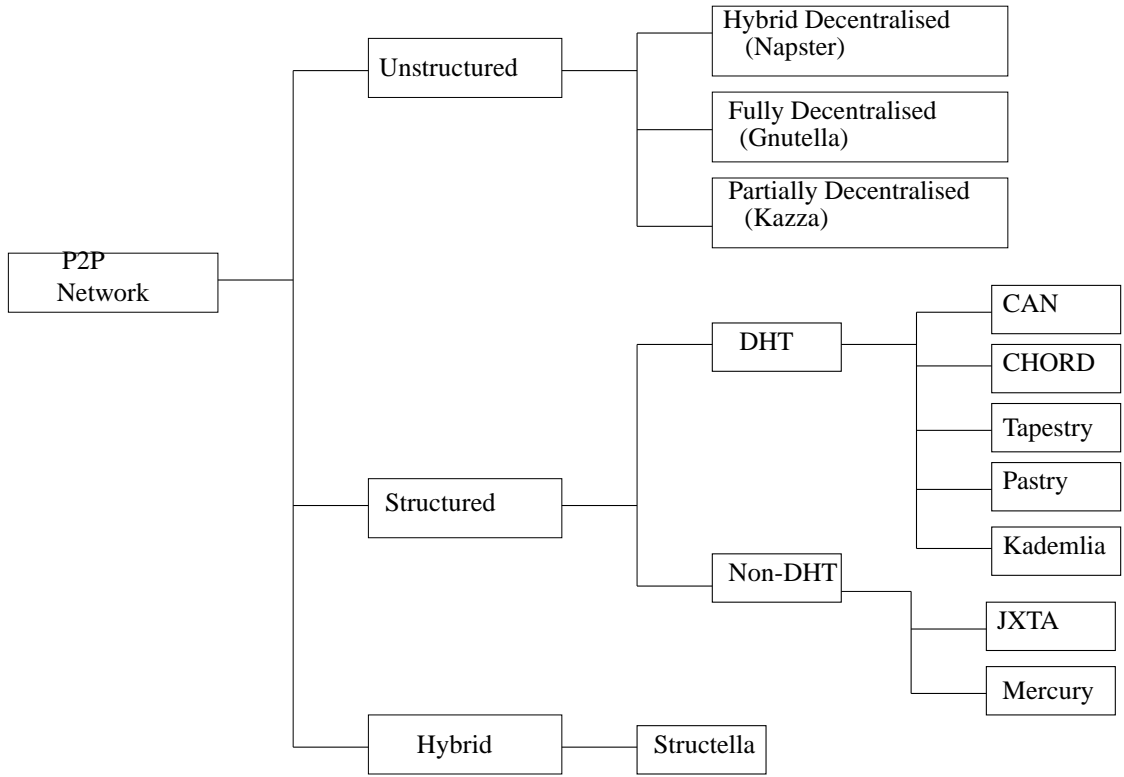

Figure 13: Peer-to-Peer network organization taxonomy

DHTs offer deterministic query search results within logarithmic bounds on network message complexity. Details about some commonly used DHTs and their various trade offs can be found in Section 4.1. Other classes of structured systems such as Mercury and JXTA do not apply randomising hash functions for organising data items and nodes. The Mercury system organises nodes into a circular overlay and places data contiguously on this ring. As Mercury does not apply hash functions, data partitioning among nodes is non-uniform. Hence it requires an explicit load-balancing scheme. It can be argued that the JXTA system is an unstructured system as it relies on search hubs for look-up operation ( similar to super-peer concept in Kazza system). However, the JXTA system offers deterministic search performance so we classify it as a non-DHT based structured P2P system.

In recent developments, new generation P2P systems have evolved to combine both unstructured and structured P2P networks. We refer to this class of systems as hybrid. Structella [34] is one such P2P system that replaces the random graph model of an unstructured overlay (Gnutella) with a structured overlay, while still adopting the search and content placement mechanism of unstructured overlays to support complex queries. In Table 4, we summarize the different P2P routing substrate that are utilized by the existing algorithms for organizing a GRIS.

### 4.3 Data Organisation Taxonomy

Traditionally, DHTs are efficient for single dimensional queries such as finding all resources that match the given attribute value. Extending DHTs to support Multi-dimensional Range Queries (MRQ), to index all resources whose attribute value overlap a given search space, is a complex problem. MRQs are based on ranges of values for attributes rather than on specific values. Compared to single-dimensional queries, resolving MRQs is far more complicated, as there is no obvious total ordering on the points in the attribute space. Further, the query interval has varying size, aspect ratio and position such as a window query. The main challenges involved in adapting MRQs in a DHT network [55] include efficient: (i) data distribution mechanisms; and (ii) data indexing or query routing techniques. In this section, we discuss various data distribution mechanisms while we analyse data indexing techniques in the next section.

A data distribution mechanism partitions the multidimensional attribute space over the set of peers in a DHT network. Efficiency of the distribution mechanism directly governs how the query processing load is distributed among the peers. A good distribution mechanism shall possess the following characteristics [55]: (i) locality-Tuples or data points nearby in the attribute space should be mapped to the same node, hence limiting the lookup complexity; (ii) load balance-number of data points indexed by each peer should be approximately the same to ensure uniform distribution of query processing [27, 70]; and (iii) minimal metadata-Prior information required for mapping the attribute space to the peer space. In decentralised settings, where peers join and leave dynamically, update policies such as the transfer of data points to a newly joined peer, should cause minimal network traffic. In the current P2P literature (refer to section 5), multidimensional data distribution mechanisms based on the following structures have been proposed (refer to Fig. 5): (i) space filling curves; (ii) tree-based structures; and (iii) variant of SHA-1/2 hashing. In Table 5, we summarise various data structures which have been applied in different algorithms to data distribution. Further, in Table 6, we present classification of the existing algorithms based on number of routing overlays utilized for managing d-dimensional data.

The Space Filling Curves (*SFCs*) [16], [67] includes the Z-curve [85] and Hilbert's curve [68]. SFCs map the given d-dimensional attribute space into a 1-dimensional space. The work in [15] utilises space-filling curves (SFC). In particular the reverse hilbert SFC is used for mapping a 1-dimensional attribute space to a two-dimensional CAN P2P space. Similarly, the work in [100] uses hilbert SFC to map an d-dimensional index space into a 1-dimensional space. The resulting 1-dimensional indexes are contiguously mapped on the Chord P2P network. The approach proposed in [55] utilises Z-curves for mapping d-dimensional space to 1-dimensional space. SFCs exhibit the locality property by mapping the points that are close in d-dimensional space to adjacent spaces in the 1-dimensional space. However, as the number of dimensions increases, locality becomes worse since SFCs suffer from "curse of dimensionality". Further, SFC based mapping fails to uniformly distribute the load among peers if the data distribution is skewed. Hence, this leads a to non-uniform query processing load for peers in the network.

Some of the recent works [108, 40, 56, 89] utilize tree-based data structures for organising the data. The approach proposed in [108] adopts the MX-CIF quadtree index to P2P networks. A distributed quadtree index assigns regions of space (a quadtree block) to the peers. In case the query processing load at a peer exceeds a predetermined threshold, then recursive subdivision of the quadtree block can be performed. The newly generated blocks are then randomly mapped to the peers. Hence, this approach has better load balancing properties as compared to SFC-based approaches in the case of a skewed data set. Other approach proposed in the work [55] utilises K-D tree based data-structure for maintaining a d-dimensional on a P2P network. However, this approach fails to answer the load balancing issue in cases where the data distribution is non-uniform.

Other approaches including [109, 31] manipulates existing SHA-1/2 hashing for mapping d-dimensional data to the peers. MAAN addresses the single dimensional range query problem by mapping attribute values to the Chord identifier space via a uniform locality preserving hashing scheme. A similar approach is also utilized in [111]. However, this approach shows poor load balancing characteristics when the attribute values are skewed.

### 4.4 Data Indexing or Query Routing Taxonomy

Fig.15 shows the query routing taxonomy for P2P systems. The query routing can be either classified as *deterministic* or a *non-deterministic* [76]. Deterministic routing means that the look-up operation will be successful within predefined bounds. Deterministic search performance can be guaranteed using either a centralised index or a distributed index in the system. Under centralised organisation, a specialised (index) server maintains the indexes of all objects in the system (e.g Napster, BitTorrent). The resource queries are routed to index servers to identify the peers currently responsible for storing the desired object. The index server can obtain the indexes from peers in one of the following ways: (i) peers directly inform the server about the files they are currently holding (e.g. Napster); or (ii) by crawling the P2P network (approach similar to web search engine). The look up operations of files in these systems are resolved with complexity $O(1)$. In distributed index organisation, every peer in the network maintains indexes for the subset of objects. Peers in DHTs such as Chord, CAN, Pastry and Tapestry maintain an index for $\log(n)$ peers where $n$ is the total number of peers in the system. Object look-up in these systems is guaranteed within the logarithmic bounds. General details about routing in some common DHT networks can be found in Section 4.1.

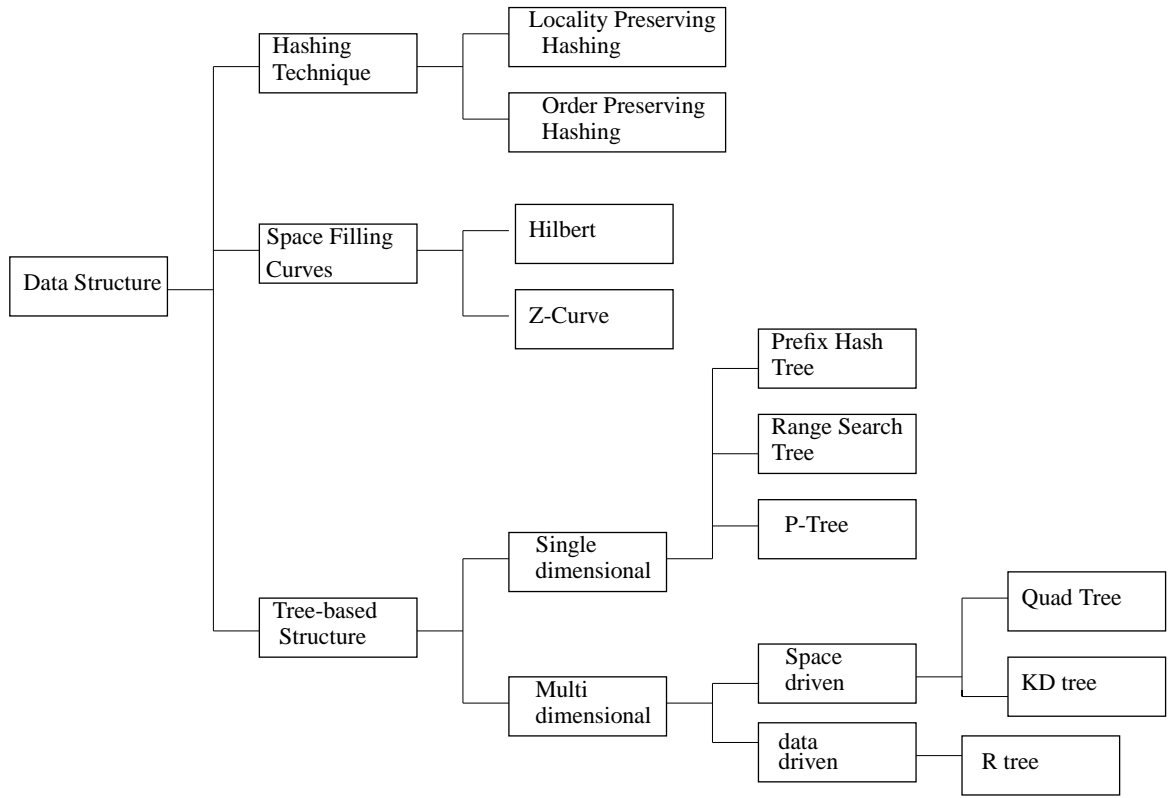

Figure 14: Data structure taxonomy

P2P network models including Gnutella, Freenet, FastTrack and Kazza offer non-deterministic query performance. Like DHTs, they maintain distributed indexes for data in the system. The Gnutella system employs query flooding model for routing object queries. Every request for objects is flooded (broadcasted) to the directly connected peers, which in turn flood their neighboring peers. This approach is used in the GRIS model proposed by the work [66]. Every RLQ message has a TTL field associated with it (i.e. maximum number of flooding hops/steps allowed). Drawbacks for flood-based routing includes high network communication overhead and non-scalability. This issue is addressed to an extent in FasTrack and Kazza by introducing the notion of super-peers. This approach reduces network overhead but still uses a flooding protocol to contact super-peers.

DHTs guarantee deterministic query lookup with logarithmic bounds on network message cost for single-dimensional queries. However, Grid RLQs are normally Multi-dimensional Range Queries (MRQ). Hence, existing routing methodologies need to be augmented with new techniques in order to efficiently resolve MRQ. In this context, various approaches have proposed different routing/indexing heuristics for resolving MRQ. Efficient query routing algorithm should exhibit following characteristics [55]: (i) routing load balance-every peer in the network on the average should route forward/route approximately same number of query messages; and (ii) low per-node state-each peer should maintain a small number of routing links hence limiting new peer join and peer state update cost. In Table 5, we summarize the query look-up complexity involved with the existing algorithms.

How an MRQ request is routed over a DHT network is directly governed by the data distribution mechanism (ref to previous section). Resolving an MRQ for a DHT network that utilises SFCs for data distribution consists of two basic steps [100]: (i) mapping the MRQ onto the set of relevant clusters of SFC-based index space; and (ii) routing the message to all peers that fall under the computed SFC-based index space. The simulation based study proposed in [55] has shown that SFCs (Z-curves) incur constant routing costs irrespective of the dimensionality of the attribute space. Routing using this approach is based on a skip graph, where each peer maintains $\log(n)$ additional routing links in the list. However, this approach has serious load balancing problems that need to be fixed using external techniques [54].

Routing MRQs in DHT networks that employ tree-based structures for data distribution requires routing to start from the root node. However, the root presents a single point of failure and load imbalance for the root peer. To overcome this, the authors in [108] introduced the concept of fundamental minimum level. This means that all the query processing and the data storage should start at that minimal level of the tree rather than at the root. Another approach [55] utilises a P2P version of kd-tree [24] for mapping d-dimensional data onto CAN P2P space. The routing utilises the neighboring cells of the data structure. The nodes in this network that manage a dense region of space are likely to have large number of neighbors, hence leading to an unbalanced routing load.

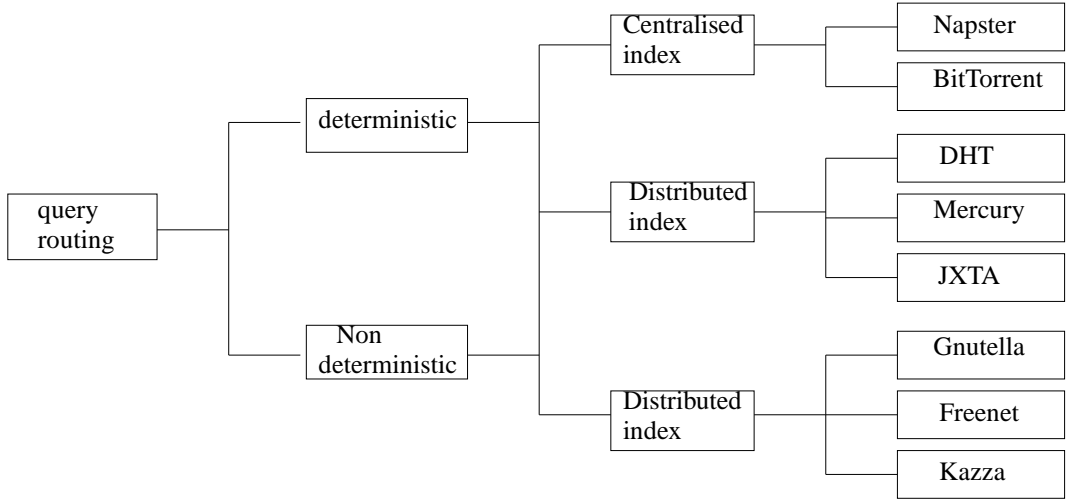

Figure 15: Query routing taxonomy

Other approaches based on variants of standard hashing schemes (such as MAAN) apply different heuristics for resolving range queries. The single-attribute dominated query routing (SAQDR) heuristic abstracts resource attributes into two categories: (i) dominant attribute; and (ii) non-dominant attribute. The underlying system queries for the node that maintains the index information for the dominant attribute. Once such a node is found, the node searches its local index information looking at satisfying the values for other non-dominant attributes in the MRQ. The request is then forwarded to the next node which indexes the subsequent range value for the dominant attribute. This approach comprehensively reduces the number of routing steps needed to resolve a MRQ. However, this approach suffers from routing load-imbalance in the case of a skewed attribute space. In Table 7, we present the classification of the existing algorithms based on query resolution heuristic, query load balancing and data locality preserving characteristic.

Table 4: Classification based on P2P routing substrate

| Routing Substrate | Network Organisation | Distributed indexing scheme name |
|---|---|---|
| Chord | Structured | PHT [89], MAAN [31], Dgrid [109], Adaptive [56], DragonFly [35], QuadTree [108], Pub/Sub-2 [111], P-Tree [40] |
| Pastry | Structured | Xenosearch [104], AdeepGrid [37], Pub/Sub-1 [107] |
| CAN | Structured | HP-Protocol [15], Flexible [100], kd-tree [55], Meghdoot [62], Z-curve [55] |
| Bamboo | Structured | SWORD [84] |
| Others | Unstructured | Mercury [26], JXTA search [58], P2PR-tree [81] |

Table 5: Classification based on data structure applied for enabling ranged search and look-up complexity

| System Name | Data Structure | Lookup Complexity |
|---|---|---|
| PHT [89] | Trie | $O(\log |D|)$, for each dimension |
| MAAN [31] | Locality preserving hashing | $O(n \times \log n + n \times s_{min})$, $s_{min}$ is minimum range selectivity per dimension; $n$ total peers |
| Dgrid [109] | SHA-1 hashing | $O(\log Y)$ for each dimension, $Y$ is the total resource type in the system |
| SWORD [84] | N.A. | N.A. |
| JXTA search [58] | RDBMS | N.A. |
| DragonFly [35] | QuadTree | N.A. |
| QuadTree [108] | QuadTree | $O(\log n + f_{max} - f_{min})$; $n$ total peers in the network; $f_{max}$ fundamental minimum level, $f_{min}$ fundamental minimum level |
| Pub/Sub-2 [111] | Order preserving hashing | $O(\log n)$; $n$ total peers, for each dimension |
| P-Tree [40] | Distributed b-+ tree | $O(m + \log_d n)$, $n$ is total peers |
| Pub/Sub-1 [107] | SHA-1 hashing | $1/2 \times O(\log n)$; Equality query, $n$ total peers, $1/2 \times O(n \log n)$, $n$ is step factor; for ranged query, in single dimension |
| Xenosearch [104] | SHA-1 hashing | N.A. |
| AdeepGrid [37] | SHA-1 hashing | N.A. |
| HP-Protocol [15] | Reverse hilbert space filling curves | N.A. |
| Flexible [100] | Hilbert space filling curve | $n \times O(\log n)$; $n$ total peers |
| Mercury [26] | N.A. | $O((\log n)/k)$; $k$ Long distance links; $n$ total peers, for single dimension |
| Adaptive [56] | Range search tree | $O(\log R_q)$; $R_q$ is range selectivity, for single dimension |
| kd-tree [55] | Kd-tree, skip pointer based on skip graphs | N.A. |
| Meghdoot [62] | SHA-1 hashing | N.A. |
| Z-curve [55] | Z-curves, skip pointer based on skip graphs | N.A. |
| P2PR-tree [81] | Distributed R-tree | N.A. |

Table 6: Classification based on No. of routing overlays for multidimensional search space

| Single | Multiple |
|---|---|
| JXTA search [58], DragonFly [35], Flexible [100], kd-tree [55],Meghdoot [62],Z-curve [55], QuadTree [108], P2PR-tree [81], Dgrid [109], SWORD [84], AdeepGrid [37], | PHT [89], MAAN [31], Adaptive [56], Pub/Sub-2 [111], P-Tree [40],Xenosearch [104], Pub/Sub-1 [107], Mercury [26],HPPROTOCOL [15] |

Table 7: Classification based on query resolution heuristic, query load balancing and data locality preserving characteristic

| System Name | Heuristic Name | Query load balancing | preserves data locality |
|---|---|---|---|
| PHT [89] | Chord routing | Good | N.A. |
| MAAN [31] | Iterative resolution, single attribute dominated routing based on Chord | Poor | N.A. |
| Dgrid [109] | Chord routing | Good | N.A. |
| SWORD [84] | Bamboo routing | N.A. | N.A. |
| JXTA search [58] | Broadcast | N.A. | N.A. |
| DragonFly [35] | Generic DHT routing | Good | No |
| QuadTree [108] | Generic DHT routing | Good | No |
| Pub/Sub-2 [111] | Chord routing | Poor | N.A. |
| P-Tree [40] | Generic DHT routing | Good | N.A. |
| Pub/Sub-1 [107] | Pastry routing | Poor | N.A. |
| Xenosearch [104] | Generic DHT routing | Poor | N.A. |
| AdeepGrid [37] | Single shot, recursive and parallel searching based on Pastry | Poor | No |
| HP-Protocol [15] | Brute force, controlled flooding, directed controlled flooding based on CAN | Poor | N.A. |
| Flexible [100] | Generic DHT routing | Poor | Yes |
| Mercury [26] | Range-selectivity based routing | Good | N.A. |
| Adaptive [56] | Generic DHT routing | Good | N.A. |
| kd-tree [55] | Skip pointer based routing | Poor | Yes |
| Meghdoot [62] | CAN based routing | Poor | Yes |
| Z-curve [55] | Skip pointer based routing | Poor | Yes |
| P2PR-tree [81] | Block/group/subgroup pointer based routing | Poor | Yes |

# 5 Survey of P2P based GRIS

## 5.1 Pastry Based Approaches

### 5.1.1 Pub/Sub-1: Building Content-Based Publish/Subscribe Systems with Distributed Hash Tables

The content-based Publish/Subscribe (Pub/Sub) system [107] is built using a DHT routing substrate. They use the topic-based Scribe [33] system which is implemented using Pastry [96]. The model defines a unique publication and subscription message format, referred to as the *schema* for each application domain (such as a stock market or an auction market). The system is capable of handling multiple domain schemes simultaneously. Each schema includes several tables, each with a standard name. Each table maintains information about a set of attributes, including their type, name, and constraints on possible values. Further, there is a set of *indices* defined on a table, where each index is an ordered collection of strategically selected attributes. The model requires application designers to manually specify the domain schemes.

When a request (publication or subscription) is submitted to the system, it is parsed for various index digests. An index digest is a string of characters that is formed by concatenating the attribute type, name, and value of each attribute in the index. An example index digest is $[USD : Price : 100 : Inch : Monitor : 19 : String : Quality : Used]$. Handling publication/subscription with exact attribute values is straightforward as it involves hashing the published request or subscription request. When a publication with attribute values that match the subscription is submitted to the system, it is mapped to the same hash key as the original subscription and forwarded to the desired node in the network. The model optimizes the processing of popular subscription (many nodes subscribing for an event) by building a multicast tree of nodes with the same subscription interest. The root of the tree is the hash key's home node (node at which publication and subscription request is stored in the network), and its branches are formed along the routes from the subscriber nodes to the root.

The system handles range values by building a separate index hash key for every attribute value in the specified range. This method has serious scalability issues. The proposed approach to overcome this limitation is to divide the range of values into intervals and a separate hash key is built for each such index digest representing that interval. However, this approach can only handle range values of single attribute in a index digest (does not support multi-attribute range value in a single index digest).

### 5.1.2 XenoSearch: Distributed Resource Discovery in the XenoServer Open Platform

XenoSearch [104] is a resource discovery system built for XenoServer [63] execution platform. The XenoServer system supports distributed resource sharing and allows users to run programs at points throughout the network to reduce communication latency, avoid network bottlenecks and minimize long-haul network changes. Xenosearch indexes resource information advertised periodically by XenoServers. An advertisement contains information on the identity, ownership and location of the XenoServer, its total and available resources and their prices. The XenoSearch system converts the advertisements to points in an $n$-dimensional space, different dimensions representing different attributes including topological location and QoS attributes. Note that, each dimension is stored in a separate DHT. The XenoSearch system is built over the Pastry [96] overlay routing protocol. This implies various XenoSearch servers are organized as nodes in the Pastry overlay network.

A separate Pastry ring operates for each dimension with XenoSearch nodes registering separately in each ring. A XenoServer registers for each dimension and derives a $key$ from its co-ordinate position in the dimension. The $key$ for a Xenoserver is therefore likely to be different in each ring. In each dimension, the resource information is logically held in the form of a tree where the leaves are the individual XenoServers and interior nodes are $aggregation\ points$ (APs) which summarizes the membership of ranges of nodes below them. These $APs$ are identified by locations in the key space which can be determined algorithmically by forming keys with successively longer suffixes. The XenoSearch node closest in the key space to an $AP$ is responsible for managing this information and for dealing with messages it receives as a consequence. This locality is provided by the proximity-aware routing characteristic of the Pastry system. Multi-dimensional range searches are performed by making a series of search requests for each dimension and finally computing their intersection.

### 5.1.3 AdeepGrid: Peer-to-Peer discovery of computational resources for grid applications

AdeepGrid [37] presents an algorithm for Grid resource indexing based on the Pastry DHT. The proposed GRIS model incorporates both static and dynamic resource attributes. The multidimensional attribute space (with static and dynamic attributes) is mapped to the DHT network by hashing the attributes. The resulting key forms the Resource ID and is also the key for the Pastry ring. The key size in the proposed scheme is 160-bit long as compared to 128-bit in Pastry. In this case the first 128-bits encode the static attributes while the remaining 32-bits encode the dynamic attributes.

The static part of the ResourceID is mapped to a fixed point while the dynamic part is represented by potentially overlapping arcs on the overlay. The beginning of each arc represents a resource's static attribute set and the length of the arc signifies the spectrum of the dynamic states that a resource can exhibit. In this case, the circular node Id space effectively contains only a finite number of nodes while they store an infinite number of objects representing dynamic attributes. RUQ are periodically initiated if the dynamic attribute value changes by a significant amount (controlled by a system-wide UCHANGE parameter). Such updates are carried out using an UPDATE message primitive. However, in some cases the new update message may map to a different node (due to a change in an attribute value) as compared to a previous INSERT or UPDATE. This can lead to defunct objects in the system. The proposed approach avoids this by having nodes periodically flush resource entries that have not changed recently or by sending REMOVE messages to prior node mappings.

Resolving RLQ involves locating the node that currently hosts the desired resource attributes (ResourceID). This is accomplished by utilizing standard Pastry routing. Three different heuristics for resolving the RLQs are proposed: (i) single-shot searching; (ii) recursive searching; and (iii) parallel searching. Single shot searching is applied in cases where the Grid application implements local strategies for searching. In this case a query for a particular kind of resource is made and if the search was successful then the node hosting the desired information replies with a REPLY (that contains resource information) message. On the other hand, recursive searching is a TTL (time to live) restricted search that continuously queries the nodes that are likely to know the desired resource information. At each step the query parameters, in particular the dynamic attribute search bits are tuned. Such a tuning can help to locate resources that may not match exactly, but that are close approximations of the original requirements. Finally, the parallel search technique initiates multiple search queries in addition to a basic search for the exact match requested parameters. This approach significantly cuts down on the query response time.

However, the above approach has the following shortcomings: (i) poor load balancing in the case of skewed data sets; and (ii) high network insert, lookup and update cost as each distinct value for a dynamic attribute creates a new key which is mapped to a different peer.

## 5.2 Chord Based Approaches

### 5.2.1 DGRID: A DHT-Based Grid Resource Indexing and Discovery Scheme

Work by Teo et al. [109] proposed a model for supporting GRIS over the Chord DHT. The unique characteristic about this approach is that the resource information is maintained in the originating domain. It leverages existing data-indexing of the Chord DHT for resolving multi-attribute queries. DGRID makes the following assumptions: (i) the total number of compute-resource types in a computational Grid is small, i.e. thousands or less, and these resource types significantly overlap across different domains; and (ii) the compute-resource types available per domain is smaller as compared to the total number of compute-resource types in all domains.

Every domain in DGRID designates an index server to the CHORD network. The index server maintains state and attribute information for the local resource. The model distributes the multi-attribute resource information over the overlay using the following scheme: a computational Grid $G = \{d\}$, where $d$ represents an administrative domain. Every domain $d = \{S, R, T\}$, consists of $S$ which is an index server such as MDS [50], $R$ is a set of compute resources, and $T = \{a\}$ is a set of resource types, where $a = \{attr\_type, attr\_value\}$ (e.g. $\{CPUSpeed, 1.7GHz\}$). An index server $S$ maintains indices to all the resource types in its domain, $S = \{r\}$. An index $r$ is defined as $r = \{t, d\}$, which denotes that $r$ is a pointer to resource type $t$. There is a one-to-one relationship between $S$ and $T$. The indices $r$ are also referred to as $data$ where key is the resource type $t$. Each data is assigned an identifier based on its key such that $id(r) = id(t)$. Conceptually, this index server is virtualized to several nodes (virtual servers) subject to the number of resource types about which the index server maintains information. The domain specific nodes are arranged as an structured overlay network (Chord nodes). Thus, DGRID virtualizes an index server $S$ into $T$ nodes in the same domain $d$. Effectively, each node represents one index $r \in S$ for domain $d$.

DGRID is based on the following axioms: (i) data is stored in its own domain, i.e. $r = (t, d)$ must be stored on $S$, the index server of domain $d$; (ii) users must be allowed to perform a look-up operation $r$ without specifying the domain name $d$. DGRID avoids node identifier collisions by splitting it into two parts: a prefix that denotes a data identifier $r$ and a suffix that denotes an index-server identifier $S$. Given node $n$ representing $r = (t, d)$, the $m$-bit identifier of $n$ is the combination of $i$-bit identifier of t, where $i \leq m$, and $m - i$ bit identifier of $S$. So effectively, $id_m(n) = id_i(t) \oplus id_{m-i}(S)$. Hence, DGRID guarantees that all $id_m(n)$ are unique, given that the identifiers of two nodes differ in either prefixes or suffixes.

The system initialization process requires the index server $S$ to perform the virtualization of its indices onto $T$ virtual servers. Each virtual server joins the DGRID system to become an overlay Chord node. This process is referred to as a $join$. The search or look-up operation in DGRID is based on Chord look-up primitives. A given key $p$, is mapped to a particular virtual index server on the overlay network using the query $get(p)$. DGRID routes each query to the node that holds the required resource information. The DGRID overlay network also supports domain specific resource type search. To facilitate such look-up requests, the index $S$ of domain $d$ is identified by $id'_{m-i}(S) = id_j(d) \oplus id_{m-i-j}(S), j < (m - i)$. In this case the query for resource $n$ of type $t$ is routed to a node $n$ that maps $S$ where $prefix_j(id'_{m-i}(S)) = id_j(d), d \in D$.

In general, a query $q$ to look-up a resource type $t$ is translated to query $q'$, $id_m(q') = id_i(t) \oplus 0$. This is done as $id(t)$ is $i$-bit length, whereas the identifier space is $m$-bit long. The look-up cost is bounded by the underlying Chord protocol i.e. $O(\log N)$. The look-up operation supports several optimizations including: (i) query $q'$ reply from all the nodes that share $prefix_i(id_m(n)) = id_i(t)$; (ii) routing within the same domain virtual servers if that resource type is hosted in the domain (this efficiently decreases the overall look-up complexity); (iii) when deciding the next routing hop for query $q'$, the node $n$ checks if its routing table contains entry for node $n'$ such that they share the same prefix. In this case, the message is routed to the node $n'$ otherwise the underlying Chord routing algorithm is applied.

The complexity involved with the node join process is $O(\log^2 N)$. The cost of index server setup is $O(g \log^2 N)$. Each index server maintains $O(g \log N)$ fingers, as it is virtualized to $g$ nodes and each node maintains $O(\log N)$ fingers. In general the look-up cost for a particular resource type $t$ is $O(\log Y)$, $Y$ is the total number of resource types available in the network. However, this approach has the following shortcomings: (i) MRQ queries needs to be split into a large number of single attribute exact match queries, hence causing a large network overhead; and (ii) external heuristics are required to map MRQ queries to same domain i.e. to a single resource that satisfies all values for the query.

### 5.2.2 Adaptive: An Adaptive Protocol for Efficient Support of Range Queries in DHT-based Systems

The work in [56] presents an algorithm to support range queries based on a distributed logical Range Search Tree (RST). Inherently, the RST is a complete and balanced binary tree with each level corresponding to a different data partitioning granularity. The system abstracts the data being registered and searched in the network as a set of attribute-value pairs (AV-pairs): $\{a_1 = v_1, a_2 = v_2, \ldots, a_n = v_n\}$. It utilizes the well known Chord DHT for distributed routing and network management issues. A typical range query with range length $R_q$ is resolved by decomposing it into $\log(R_q)$ sub-queries. These sub-queries are sent to the nodes that index the corresponding data. The system supports updates and queries for both static and dynamic resource attributes.

The content represented by AV-pair is registered with the node whose $ID$ is numerically closest to the hash of the AV-pair in the name. To overcome the skewed distribution of AV-pairs over nodes in the DHT, the system organizes nodes in a logical load balancing matrix (LBM). Each column in the LBM represents a partition, i.e. a subset of content names that contain a particular AV-pair, while nodes in different rows within a column are a replica of each other. Initially, a LBM has only one node but whenever the registration load on a particular node in the system exceeds a threshold ($T_{reg}$) then the matrix size is increased by 1. All future registration requests are shared by the new nodes in the LBM. Note that, the number of partitions $P$, is proportional to the registration load: $P = \lceil \frac{L^R}{C_R} \rceil$, where $L^R$ is the data item's registration load, and $C_R$ is the capacity of each node.

An attribute $a$, can have numerical values denoted by domain $D_a$. $D_a$ is bounded and can be discrete or continuous. $D_a$ is split up into sub-ranges and assigned to different levels of the RST. An RST with $n$ nodes has $\lceil \log n + 1 \rceil$ levels. Levels are labeled consecutively with the leaf level being level 0. Each node in the RST holds indexing information for different sub-ranges. Typically, the range of the $i$-th node from the left represents the range $[v_i, v_{i+2^l-1}]$. The union of all the ranges at each level covers the full $D_a$. In a static RST, the attribute value $v$ is registered at each node in the tree which lies on the path $path(v)$ to the leaf node that indexes the exact value. The new value information is updated into the LBM if a node on the path maintains it.

In a static setting, a query $Q : [s, e]$ for values of an attribute $a$ is decomposed into k sub-queries, corresponding to $k$ nodes in the RST, $N_1, \ldots, N_k$. The efficiency of the query resolution algorithm depends on the $relevance$ factor which is given by $r = \frac{R_q}{\sum_{i=1}^{k} R_i}$, where $R_i$ is node $N_i$'s range length, and $R_q$ is the query length. The relevance factor $r$ denotes how efficiently the query range matches the RST nodes that are being queried. The query $Q$ is resolved by querying the node which has the largest range within $[s, e]$ (also referred to as the node which has the minimum cover (MC) for the query range). Furthermore, this process is recursively repeated for the segments of the range that are not yet decomposed. When the MC is determined, the query is triggered on all the overlay nodes that correspond to each MC node.

Note that, in a static setting the attribute value registration information goes to every level of the RST. In the dynamic range query design, the authors propose that an attribute value should be registered only with the nodes that are needed based on the query ranges and the load information. Hence this minimizes the number of messages and load on the root node in the RST. These set of nodes are called $band$ nodes and only these nodes accept registrations and query messages. To enable such a constrained search and register mechanism the system applies various protocols including the Path Maintenance Protocol (PMP). The PMP propagates information to nodes in the RST about the band for registrations and queries. This information is referred to as the Path Information Base (PIB). More details about the query resolution and registration in dynamic settings based on band information can be found [56].

### 5.2.3 Pub/Sub-2: Content-based Publish-Subscribe Over Structured P2P Networks

The work in [111] presents a content-based publish-subscribe system by leveraging the routing and network management functionality of the Chord DHT. More importantly, the proposed scheme supports publication and subscription matching involving range predicates (attribute values lie over a range). The model defines the event schema which is a set of typed attributes. Every attribute $a_i$ consists of a type, a name and a numeric or string value $v(a_i)$. The attribute type belongs to a predefined set of primitive data types commonly defined in most programming languages. The attribute name is normally a string, whereas the value can be a string or numeric in any range defined by the minimum and maximum $(v_{min}(a_i), v_{max}(a_i))$ value along with the attribute's precision $v_{pr}(a_i)$. The model supports a generalized subscription schema that includes a rich set of subscriptions which may contain different data types such as integers, strings etc, and common operators such as $=, \neq, >, <$. A notification event matches a desired subscription if and only if all the constraints defined on the subscription's attributes are satisfied. With every subscription, the model associates a unique Subscription Identifier (subID). The subID is the concatenation of three parts- $c_1$, $c_2$ and $c_3$. $c_1$ is the $id$ of the node which is receiving the subscription, in this case it is $m$-bits in a Chord ring with an $m$-bit identifier address space. $c_2$ is the $id$ of the subscription itself, and $c_3$ is the number of attributes on which the constraints are declared.

The model maintains a Chord ring for every unique attribute in a subscription (i.e. if a subscription involves $n$ attributes then there would be $n$ separate Chord rings). An attribute $a_i$ of a subscription with identifier $subID$ is placed on a node $successor(h(v(a_i)))$ in the Chord ring. A subscription can declare a range of values for the attribute, $a_i$, such as $v_{low}(a_i)$ and $v_{high}(a_i)$. In this case, the model follows $n$ steps, where $n = \frac{v_{high}(a_i) - v_{low}(a_i)}{v_{pr}(a_i)}$, at each step a Chord node is chosen by the $successor(h(v_{low}(a_i) + v_{pr}(a_i)))$ function. In the subsequent steps the previous attribute value is incremented by the precision value $v_{pr}(a_i)$ and mapped to the corresponding Chord node. Updating the range values is done by following the same procedure for all Chord nodes that store the subID for the given range of values. If the new range is different from the old ones then Chord nodes are added or removed from the attribute specific overlay network. The overall message routing complexity depends on the type of constraints defined over the attributes for a given $subID$. In case of equality constraints, the average number of routing hops is $1/2 \log(n)$. When the constraint is a range then the complexity involved is $n \times 1/2 \log(n)$, where $n$ is the step factor.

An information publish event in the system is denoted by $N_{a-event}$ that includes various attributes with values. The event-publish to event-notify matching algorithm processes each attribute associated with $N_{a-event}$ separately. It locates various nodes that store the subIDs for an attribute $a_i$ for the value $v(a_i)$ applying the function $successor(h(v(a_i)))$. The matching algorithm then stores the list of unique subIDs, that are found at a node $n$ in the list $L_{a_i}$ designated for $a_i$. The $N_{k-sub}$ list stores the subIDs that match the event $N_{a-event}$. A $subID_k$ matches an event if and only if it appears in exactly $N_{k-sub}$ derived from the Chord ring. Note that, parsing the $subID_k$ field gives the IP address of the node which triggered the subscription. Hence, the notification message is delivered directly to the underlying nodes using the address. The overall message routing complexity involved in locating the list of subIDs matching an event $N_{a-event}$ is $1/2 \log(n)$.

The model also proposes the routing optimization algorithm by applying the order preserving Chord for range attributes. In this case the lowest attribute value $v_{low}(a_i)$ is mapped to a node in the network (with complexity $1/2 \log(n)$), and subsequent range values in $r$ steps are assigned to successive nodes in the overlay network. This approach leads to $r + 1/2 \log(n)$ hops in total.

### 5.2.4 QuadTree: Using a Distributed Quadtree Index in the Peer-to-Peer Networks

The work in [108] proposes a distributed quad-tree index that adopts an MX-CIF quadtree [98] for accessing spatial data or objects in P2P networks. A spatial object is an object with extents in a multidimensional setting. A query that seeks all the objects that are contained in or overlap a particular spatial region is called a spatial query. Such queries are resolved by recursively subdividing the underlying multidimensional space and then solving a possibly simpler intersection problem. This recursive subdivision process utilizes the basic quad-tree representation. In general, the term quad-tree is used to describe a class of hierarchical data structures whose common property is that they are based on the principle of common decomposition of space.

The work builds upon the region quad-tree data structure. In this case, by applying the fundamental quad-tree decomposition property the underlying two-dimensional square space is recursively decomposed into four congruent blocks until each block is contained in one of the objects in its entirety or is not contained in any of the objects. The distributed quad-tree index assigns regions of multidimensional space to the peers in a P2P system. Every quad-tree block is uniquely identified by its centroid, termed as the control point. Using the control point, a quad-tree block is hashed to a peer in the network. The Chord method is used for hashing the blocks to the peers in the network. If a peer is assigned a quad-tree block, then it is responsible for processing all query computations that intersects the block. Multiple control points (i.e. quad-tree blocks) can be hashed to the same peer in the network. To avoid a single point of failure at the root level of the quad-tree the authors incorporate a technique called *fundamental minimum level,* $f_{min}$. This technique means that objects are only allowed to be stored at levels $l \geq f_{min}$ and therefore all the query processing starts at levels $l \geq f_{min}$. The scheme also proposes the concept of a *fundamental maximum level, $f_{max}$*, which limits the maximum depth of the quad-tree at which objects are inserted.

A peer initiates a new object insertion or query operation by calling the methods InsertObject() or ReceiveClientsQuery(). These methods inturn call a subdivide() method that computes the intersecting conrol point associated with the new object or look-up query. Once the control points are computed, the peer broadcasts the insertion or query operation to the peer(s) that own(s) the respective control points. The contacted peers evokes DoInsert() and DoQuery() methods to determine the location for the inserted object or to locate the peers that can answer the query. The operation may propagate down to the $f_{max}$ level or until all relevant peers are located. The total message complexity involved with inserting a new object or resolving a look-up query is $O(f_{max})$. The authors also propose some optimizations such as each node maintains a cache of addresses for its immediate children in the hierarchy. This reduces the look-up complexity to $O(1)$, as it is no longer required to traverse the Chord ring for each child. However, this is only true when the operation is a regular tree traversal.

### 5.2.5 DragonFly: A Publish-Subscribe Scheme with Load Adaptability

The work in [35] proposes a content-based publish-subscribe system with load adaptability. They apply a spatial hashing technique for assigning data to the peers in the network. The system supports multi-attribute point and range queries. The query routing and object location (subscription and publication) mechanism can be built using the services of any DHT. Each distinct attribute is assigned a dimension in a multi-dimensional Cartesian space. Hence, a domain with $n$ attributes $\{A_1, A_2, \ldots, A_n\}$ will be represented by an $n$-dimensional Cartesian space. Every attribute in the system has lower and upper bound on its values. The bounds act as constraints for subscriptions and events indexing. The $n$-dimensional Cartesian space is arranged as a tree structure with the domain space mapped to the root node of the tree. In particular, the tree structure is based on a quad tree [98]. To negate a single point of failure at the root node, system adopts a technique called the *fundamental minimum level*. More details about this technique can be found in [108]. This technique recursively divides the logical space into four quadrants. With each recursion step on a existing quadrant, four new quadrants are generated. Hence, multiple recursion steps basically create a mutli-level quad tree data structure. The quad tree based organization of DragonFly introduces parent-child relationships between tree cells. A cell at a level $n$ is always a child of a particular cell at level $n-1$. However, this relationship exists between consecutive levels only. In other words, every cell has a direct relationship with its child cells and no relationship with its grandchild cells. Another important feature of DragonFly is the diagonal hyperplane. This hyperplane is used to handle publish and subscribe region pruning and selection in $n$-dimensional space. In 2-d space, the diagonal hyperplane is a line spanning from the north-west to the south-east vertices of the rectangular space. In multi-dimensional context, this hyperplane is represented by the equation $\frac{x_1}{x_{max_1} - x_{min_1}} + \frac{x_2}{x_{max_2} - x_{min_2}} + \ldots + \frac{x_n}{x_{max_n} - x_{min_n}} = K$, where $x_{max_n}$ and $x_{min_n}$ are the upper and lower boundary values for $n$-th attribute in the domain space.

The multi-dimensional domain space acts as the basis for object routing in DragonFly. Every subscription is mapped to a particular cell or set of cells in the domain space. In this case, the cell acts as the subscription container. A point subscription takes the form $\{A_1 = 10, A_2 = 5\}$ while a range subscription is represented by $\{A_1 \leq 10, A_2 \leq 5\}$. The root cells at the fundamental minimum level are the entry points for a subscription's object routing. These root cells are managed by the peers in the network. Every subscription is mapped to a particular region in the multi-dimensional space. The peer responsible for the region (root cell) is located by hashing the coordinate values. If the root cell has undergone the division process due to overload, then the child cells (peers at lower level in the hierarchy) are searched using the DHT routing method. Once a child cell is located, the root cell routes the subscription message to it. This process is repeated untill all relevant child cells are notified for this subscription. However, if the root cell has not undergone any division process then it is made responsible for this subscription.

Mapping publication events to the peers in the network is similar to the subscription mapping process. There are two kinds of publishing events i.e. point and range event. Mapping point events is straightforward, as the relevant root cell (peer) is located by hashing the coordinated values. Resolving cells corresponding to range events can be complex. In this case, the routing system sends out the published event to all the root cells that intersect with the event region. When the message reaches the root cells, a method similar to the one adopted in case of the subscription events is applied to locate the child cells.

### 5.2.6 MAAN: A Multi-Attribute Addressable Network for Grid Information Services

Cai et al. [31] present a multi-attribute addressable network (MAAN) approach for enabling GRIS. They extend the Chord [106] protocol to support MRQs. MAAN addresses the range query problem by mapping the attribute values to the Chord identifier space via uniform locality preserving hashing. It uses SHA-1 hashing to assign an $m$-bit identifier to each node and to each attribute value of the string type. For attributes with the numerical values MAAN applies locality preserving hashing functions to assign each attribute value an identifier in the $m$-bit space. A locality preserving hashing (LPH) function $H$ has the following property: $H(v_i) < H(v_j)$ iff $v_i < v_j$, and if an interval $[v_i, v_j]$ is split into $[v_i, v_k]$ and $[v_k, v_j]$, the corresponding interval $[H(v_i), H(v_j)]$ must be split into $[H(v_i), H(v_k)]$ and $[H(v_k), H(v_j)]$. Further, the LPH theorem states that if one applies the LPH function $H$ to map an attribute value $v$ to the $m$-bit circular space $[0, 2^{m-1}]$, then given a range query $[l, u]$ where $l$ and $u$ are the lower bound and upper bound respectively, in this case the nodes which maintain index for attribute value $v$ in $[l, u]$ must have an identifier equal to or larger than $successor(H(l))$ and equal to or less than $successor(H(u))$.

The basic range query by a node $n$ includes the numeric attribute values $v$ between $l$ and $u$ for attribute $a$, such that $l \leq v < u$, where $l$ and $u$ are the lower and upper bound respectively. In this case, node $n$ formulates a look-up request and uses the underlying Chord routing algorithm to route it to node $n_l$ such that $n_l = successor(H(l))$. The look-up is done using the $SEARCH\_REQUEST(k, R, X)$ primitive, $k = successor(H(l))$ is the key to look up, $R$ is the desired attribute value range $[l, u]$ and $X$ is the list of resources that has the required attributes in the desired range. A node $n_l$ after receiving the search request message, indexes its local resource list entries and augments all the matching resources to $X$. In case $n_l$ is the $successor(H(u))$ then it sends a reply message to the node $n$. Otherwise, the look-up request message is forwarded to its immediate successor until the request reaches the node $n_u$, the successor of $H(u)$. The total routing complexity involved in this case is $O(\log N + K)$, where $O(\log N)$ is the underlying Chord routing complexity and $K$ is the number of nodes between $n_l$ and $n_u$.

MAAN also supports multi-attribute query resolution by extending the above single-attribute range query routing algorithm. In this case, each resource has $M$ attributes $a_1, a_2, ..., a_m$ and corresponding attribute value pairs $< a_i, v_i >$, such that $1 \leq i \leq M$. Each resource registers its information (attribute value pairs) at a node $n_i = successor(H(v_i))$ for each attribute value $v_i$. Thus each node in the overlay network maintains the resource information in the form of $< attribute - value, resource - info >$ for different attributes. The resource look-up query in this case involves a multi-attribute range query which is a combination of sub-queries on each attribute dimension, i.e. $v_{il} \leq a_i \leq v_{iu}$ where $1 \leq i \leq M$, $v_{il}$ and $v_{iu}$ are the lower and upper bounds of the look-up query.

MAAN supports two routing algorithms to resolve multiple-attribute queries: (i) iterative query resolution (IQR); and (ii) single attribute dominated query resolution (SADQR). With IQR, a node $n$ that wishes to look-up for a resource having $M$ attributes within a given range, simply sends $M$ sub-queries on different attributes. Each sub query is resolved as the single attributed query and finally the result of each sub-query is compiled and returned to the node $n$. The search request message in this case is: $SEARCH\_REQUEST(k, a, R, X)$, where the additional parameter $a$ denotes the name of the attribute to be looked up and $k$, $R$ and $X$ have the same meaning similar to single attribute based query. To resolve $M$-attribute queries, the total routing hops involved is $O(\sum_{i=1}^{M}(\log N + K_i))$, where $K_i$ is the number of nodes that intersects the query range on attribute $a_i$. The number of routing steps can be reduced by selecting the minimum attribute selectivity parameter $s_i$ for attribute $i$. The value of $s_i$ is the ratio of query range width in identifier space to the size of the whole identifier space, i.e. $s_i = \frac{H(v_{iu}) - H((v_{il}))}{2^m}$. The overall routing complexity in this case is $O(\sum_{i=1}^{M}(\log N + N \times s_i))$

The SADQR method defines two kinds of attributes for a resource $m$ having attribute set $\{a\}$: (i) attribute $a_k \in a$ called a dominant attribute; and (ii remaining attribute are called non-dominant attributes, $\{a\}, \{a_k\}$. In this case a search request by node $n$ is : $SEARCH\_REQUEST(k, a_k, R, O, X)$. Both $k$ and $R$ have the same meaning similar to the iterative query resolution approach. Also, $a_k$ is the dominant attribute and $O$ is a list of sub-queries for all other non-dominant attributes. If a node $n_l$ is found that maintains the range information $l \in [l, u]$ for the attribute $a_k$, then the node $n_l$, searches its local index information satisfying all other sub queries in $O$ and appends them to $X$. The result is sent back to the $n$ if $n_l$ is also the $successor(H(u))$, otherwise the request is forwarded to the immediate successor $n_s$. This approach reduces the number of routing steps needed to resolve a multiple-attribute query as compared to IQR. The overall routing complexity involved in this case is $O(\log N + N \times S_{min})$, where $S_{min}$ is the minimum selectivity for all attributes. The main drawback of this approach includes: (i) non-uniform load-balancing if data set is skewed; and (ii) resolving MRQ incurs large message complexity.

### 5.2.7 Flexible: Flexible Information Discovery in Decentralized Distributed Systems

Schmidt et al. [100] propose a GRIS model that utilizes SFCs for mapping multi-dimensional attribute space to single-dimensional search space. The proposed GRIS model consists of the following main components: (i) a locality preserving mapping that maps data elements to indices; (ii) an overlay network topology; (iii) a mapping from indices to nodes in the overlay network; (iv) a load balancing mechanism and (v) a query engine for routing and efficiently resolving keyword queries using successive refinements and pruning.

Unlike other distributed look-up systems including ([31], [109]), the proposed approach does not apply consistent hashing for mapping data element identifiers to indices. All data elements are described using a sequence of keywords such as memory, CPU speed and network bandwidth. The keywords form the coordinates of a multidimensional space, while the data elements are the points in the $n$-dimensional keyword space. This mapping is accomplished using a locality-preserving mapping called Space Filling Curves (*SFC*) [16], [67] . SFCs are used to generate a 1-dimensional index space from an $n$-dimensions keyword space, where $n$ is the number of different keyword types. Any range query or query composed of keywords, partial keywords, or wild-cards, can be mapped to regions in the keyword space and the corresponding clusters in the SFC.

The Chord protocol is utilized to form the overlay network of peers. Each data element is mapped, based on its SFC-based index or key, to the first node whose identifier is equal to or follows the key in the identifier space. The peer join, leave and failure complexity is $O(\log_2^2 N)$ messages. The peers are efficiently located based on the content they store. The look-up cost is $O(\log_2 N)$ messages. However, the look-up operation involving partial queries and range queries typically requires interrogating more than one node, since the desired information is available at multiple nodes.

The look-up queries can consist of combination of a keywords, partial keywords or wildcards. The result of the query is a complete set of data elements that matches the user's query. Valid queries include (computer, network), (computer,net*) and (comp*,*). The range query consists of at least one dimension that is needed to be looked up for range values. A valid range query that indexes a combination of memory, CPU frequency and base network bandwidth is (256-512MB,*,10Mbps-*). The query resolution process consists of two steps: (i) translating the keyword query to relevant clusters of the SFC-based index space and (ii) querying the appropriate nodes in the overlay network for data-elements.

In case the query is based on whole keywords (no wild-cards), it is mapped to exactly one point in the index space. However, if the query contains partial keywords or wild-cards or it is a range query, then the query resolution schemes return a set of points (data elements) in the multidimensional keyword space that matches a set of points in the index space. Such set of points are referred to as clusters (marked by different patterns). Each cluster may contain zero, one or more data element that match the query. Such data clusters may be mapped to one or more adjacent nodes in the underlying overlay network. Further, a node may also store more than one cluster. Once the clusters associated with a query are identified, the query message is sent for each cluster. The cluster identifier is constructed using the SFC's digital causality property. This property guarantees that all the cells that form part of single cluster share the same first $i$ prefix digits.

The system also supports two load balancing algorithms in the overlay network. The first algorithm proposes exchange of information between neighboring nodes about their loads. In this case, the most loaded nodes give a part of their load to their neighbors. The cost involved in this operation at each node is $O(\log_2^2 N)$ messages. The second approach uses a virtual node concept. In this algorithm, each physical node houses multiple virtual nodes. The load at a physical node is the sum of the load of its virtual nodes. In case the load on a virtual node exceeds predefined threshold value, the virtual node is split into more virtual nodes. If the physical node is overloaded, one or more of its virtual nodes can migrate to less loaded neighbors or fingers. Note that, creation of virtual node is inherent to the Chord routing substrate.

### 5.2.8 P-Tree: Querying Peer-to-Peer Networks Using P-Trees

Crainniceanu et al. [40] propose a distributed, fault-tolerant P2P index structure called P-tree. The main idea behind the proposed scheme is to maintain parts of semi-independent $B^+-$trees at each peer. Chord is utilized as a P2P routing substrate. Every peer in the P2P network believes that the search key values are organized in a ring, with the highest value wrapping around to the lowest value. Whenever a peer constructs its search tree, the peer pretends that its search key value is the smallest value in the ring. Each peer stores and maintains only the *left-most root-to-leaf path* of its corresponding $B^+-$ tree. The remaining part of the sub-tree information is stored at a subset of other peers in the overlay network. Furthermore, each peer only stores tree nodes on the root-to-leaf path, and each node has at most $2d$ entries. In this case, the total storage requirement per peer is $O(d\ log_d N)$. The proposed approach guarantees $O(\log_d N)$ search performance for equality queries in a consistent state. Here $d$ is the order of the sub-tree and $N$ is the total number of peers in the network. Overall, in a stable system when no inserts or deletes operation is being carried out, the system provides $O(m + log_d N)$ search cost for range queries, where $m$ is the number of peers in the selected range.

The data structure for a P-tree node $p$ is a double indexed array $p.node[i][j]$, where $0 \leq i \leq p.maxLevel$ and $0 \leq j \leq p.node[i].numEnteries$, $maxLevel$ is the maximum allowed height of the P-tree and $NumEnteries$ is the number of entry allowed per node. Each entry of this 2-dimensional array is a pair (*value,peer*), which points to the $peer$ that holds the data item with the search key $value$. In order that the proposed scheme works properly, the P-tree should satisfy the four predefined properties. These properties includes the constraints on the number of entries allowed per node, left-most root-to leaf path, coverage and separation of sub-trees. The coverage property ensures that there are no gaps between the adjacent sub-trees. While the separation property ensures that the overlap between adjacent sub-trees at a level $i$ have at least $d$ non-overlapping entries at level $i-1$. This ensures that the search cost is $O(\log_d N)$.

The consistency of the P-tree nodes is maintained by two coordinating process, the ping process and the stabilization processes. These processes are active at each peer in the network. The ping process is responsible for detecting the inconsistencies in the P-tree nodes, if found then it marks it for future repair by the stabilization process. The stabilization process at a peer periodically repairs the marked inconsistencies. A search request is a range query that specifies a lower-bound ($lb$) and an upper-bound ($ub$) of an attribute. At present the proposed scheme supports single attribute range queries. The search procedure at each peer selects the farthest away pointer that does not exceed the $lb$ and forwards the query to that peer. when query resolution reaches the lowest level of the P-tree, it traverses the successor list until the the value of a peer exceeds $ub$. Finally, a $SearchDoneMessage$ is sent to the peer that originated the search request. The main shortcoming of this approach is that the system needs to maintain a separate P-tree for every resource dimension.

## 5.3 CAN Based Approaches

### 5.3.1 One torus to rule them all (Kd-tree and z-curve based indexing)

The work in [55] proposes two approaches for enabling MRQs over the CAN DHT. The multidimensional data is indexed using the well known spatial data structures: (i) z-curves ; and (ii) kd-tree. First scheme is referred to as SCRAP: Space Filling Curves with Range Partitioning. SCRAP involves two fundamental steps: (i) the d-dimensional data is first mapped to a 1-dimensional using the z-curves; and (ii) then 1-dimensional data is contiguously range partitioned across peers in the DHT space. Each peer is responsible for maintaining data in the contiguous range of values. Resolving MRQs in SCRAP network involves two basic steps: (i) mapping MRQ into SRQ using the SFCs; and (ii) routing the 1-dimensional range queries to the peers that indexes the desired look-up value. For routing query in 1-dimensional space the work proposes a scheme based on skip graph [17]. A skip graph is a circular linked list of peers, which are organized in accordance with their partition boundaries. Additionally, peers can also maintain skip pointers for faster routing. Every peer maintains skip pointers to $\log(n)$ other peers at a exponentially increasing distances from itself to the list. A SRQ query is resolved by the peer that indexes minimum value for the desired range. The message routing is done using the skip graph peer lists.

Other approach referred to as Multi-Dimensional Rectangulation with Kd-Trees (MURK). In this scheme, d-dimensional space (for instance a 2-d space) is represented as "rectangles" i.e. (hypercuboids in high dimensions), with each node maintaining one rectangle. In this case, these rectangles are used to construct a kd-tree. The leaf node in the tree are stored by the peers in the network. Routing in the network is based on the following schemes: (i) CAN DHT is used as basis for routing the MRQs ; (ii) random pointers-each peer has to maintain skip pointers to random peers in the network. This scheme provides similar query and routing efficiency as multiple realities in CAN; and (iii) space-filling skip graph-each peer maintain skip pointers to $\log(n)$ other peers at exponentially increasing distances from itself in the network. Simulation results indicate that random and skip-graph based routing outperforms the standard CAN based routing for MRQs.

### 5.3.2 Meghdoot: Content-Based Publish/Subscribe over P2P Networks

The work in [62] proposes a content-based Pub/Sub system based on CAN routing substrate. Basic models and definitions are based on the scheme proposed in the work [97]. The model defines a d-dimensional attribute space given by the set $\mathcal{S} = A_1, A_2, \ldots, A_n$. Further, each attribute value $A_i$ is denoted using the tuple Name:Type, Min, Max. Different Type includes a integer, floating point and string character. While Min and Max denotes the range over which values lie. All peers in the system use the same schema $S$.

Typically, a subscription is a conjunction of predicates over one or more attributes. Each predicate specifies a constant value or range using the operators (such as $=,\geq,\leq,\geq$ and $\leq$) for an attribute. An example subscription is given by $S = (A_1 \geq v_1) \wedge (v_2 \leq A_3 \leq v_3)$. A system consisting of $n$ attributes is always mapped to a cartesian space of $2n$ dimensions. An attribute $A_i$ with domain value $[L_i, H_i]$ corresponds to dimensions $2i - 1$ and $2i$ in a d-dimensional cartesian space. The $2n$ dimensional logical space is partitioned among the peers in the system. A subscription $S$ for $n$ attributes is mapped to the point $< l_1, h_1, l_2, h_2, \ldots, l_n, h_n >$ in the $2n$ dimensional space which is referred to as the subscription point. Pub/Sub applications submit their subscription to a randomly chosen peer $P_0$. A origin peer $P_0$ routes the subscription request to the target peer $P_t$ using the basic CAN routing scheme. A peer $P_t$ owns a point in the d-dimensional space to which a subscription $S$ maps. The overall complexity involved in routing a subscription is $O(d\ n^{1/d})$, where $n$ is the number of peers in the system and $d$ is the dimensionality of the cartesian space.

Similarly every publish event is mapped to a particular point in the d-dimensional space, also referred to as the event point/event zone. The event is then routed to the $P_t$ from the origin peer using the standard CAN routing. All the peers that own the region affected by a event are notified accordingly. Following this, all the peers in the affected region matches the new event against the previously stored subscriptions. Finally, the event is delivered to applications that have subscribed for the event.

### 5.3.3 HP-Protocol: Scalable, Efficient Range Queries for Grid Information Services

Andrejak et al. [15] extend the CAN routing substrate to support 1-dimensional range queries. They apply the SFC in particular the Hilbert Curves for mapping a 1-dimensional attribute space (such as no. of processors) to a d-dimensional CAN space. For each resource attribute/dimension a separate CAN space is required. To locate a resource based on multiple attributes, the proposed system iteratively queries for each attribute in different CAN space. Finally, the result for different attributes are concatenated similar to "join" operation in the database.

The resource information is organized in pairs (attribute-value,resource-ID), are referred to as objects. Thus, in this case there is one object per resource attribute. Hence, if a resource has $m$ attributes then there would be $m$ different *object* type. The range of an attribute lies in the interval $[0.0, 1.0]$. A subset of the servers are designated as information servers in the underlying CAN-based P2P network (for e.g. one information server per computational resource or storage resource domain). Each of them is responsible for a certain sub-interval of $[0.0, 1.0]$ of the attribute values. Such servers are called interval keeper (IK). Each computational resource server or storage server in the Grid registers its current attribute value to an IK. Each IK owns a zone in the logical $d$-dimensional Cartesian space (a d-torus).

The CAN space is partitioned into zones, with a node (in this case an information server) serving as a. Similarly, objects (in this case (attribute, value) pair) is mapped to logical points in the space. A node $R$ is responsible for all the objects that are mapped to its zone. It is assumed that the dimension $d$ and the Hilbert Curve's approximation level is 1 are fixed, and known throughout the network. Given a (attribute,value) pair, a hypercube is determined by the Hilbert Function, the function returns the corresponding interval that contains the value. Following this, the message containing this object is routed to an IK whose zone encompasses this hypercube.

Given a range query $r$ with lower and upper bounds $\in [l, u]$, a query message is routed to an information server which is responsible for the point $\frac{l+u}{2}$. Once such a server is located, then the request is recursively flooded to all its neighbors until all the IKs are located. Three different kinds of message flooding scheme are presented including the brute force, controlled flooding and directed control flooding. Each of these scheme has different search strategy and hence have different message routing complexities. The system handles server failures/dynamicity by defining an information update interval. If the update for one of the objects is not received in the next reporting round, the corresponding object is erased/removed from the network. In case, the object value changes (attribute value) to the extent that it is mapped to a new IK then previous object is erased in the next reporting round. The main shortcoming includes the system need to maintain separate CAN space for each resource dimension.

## 5.4 Miscllaneous

### 5.4.1 SWORD: Distributed Resource Discovery on PlanetLab

SWORD [84] is a decentralized resource discovery service that supports multi-attribute queries. This system is currently deployed and tested over PlanetLab [10] resource sharing infrastructure. It supports different kind of query composition including per-node characteristics such as load, physical memory, disk space and inter-node network connectivity attributes such as network latency. The model abstracts resource as a networks of interconnected resource groups with intra-group, inter-group, and per-node network communication attributes. In particular, SWORD system is a server daemon that runs on various nodes. The main modules of the daemon includes the distributed query processor (DQP) and the query optimizer (QO). SWORD system groups the nodes into two sets. One set of nodes called server nodes form the part of the structured P2P overlay network [1, 26] and are responsible for managing the distributed resource information. While other set of nodes are computation nodes that report their resource attribute values to these server nodes.

For each resource attribute $A_i$, a corresponding DHT key $k_i$ is computing using the standard SHA-1 scheme. A key $k_i$ is computed based on the corresponding value of $A_i$ at the time attribute value is sent. Each attribute is hashed to a 160-bit DHT key. The mapping function convert attribute values from their native data-type (String) and range (numeric) to a range of DHT keys. On receiving the attribute value tuple, the server node stores the tuple in the local table. In case, these values are not updated within timeout interval then are deleted (assuming node has probably left the network or owner of the key has changed due to change in attribute values). SWORD resolves multi-attribute range query similar to [26]. The system maintains a separate overlay for every attribute $A_i$.

Users in general specify resource measurements values including the node characteristics and inter/intra-node network latency. A query also includes the node characteristics such as penalty levels for selecting nodes that are within the required range but outside the preferred range. These queries are normally written in Extended Markup Language (XML). A user submits query to a local DQP which in turn issues a distributed range query. Once the result is computed, then it is passed on to the QO (the nodes in result that are referred as "candidate nodes"). The QO selects those candidate nodes which has least penalty and passes the refined list to the user.

### 5.4.2 Mercury: Supporting Scalable Multi-Attribute Range Queries

Mercury [26] is a distributed resource discovery system that supports multi-attribute based information search. Mercury handles multi-attribute lookups by creating a separate routing hub for every resource dimension. Each routing hub represents a logical collection of nodes in the system and is responsible for maintaining range values for a particular dimension. Thus, hubs are basically orthogonal dimensions in the d-dimensional attribute space. Further, each hub is part of a circular overlay network. Mercury system abstracts the set of attributes associated with an application by $\mathcal{A}$. $\mathcal{A}_\mathcal{Q}$ and denotes the set of attributes in a query message using $\mathcal{Q}$. Attribute set for data-record $\mathcal{D}$ is denoted by $\mathcal{A}_\mathcal{D}$. The function $\pi_a$ returns the value (range) for a particular attribute $a$ in a query. A attribute hub for an attribute $a$ is denoted by $H_a$. Each node in a $H_a$ is responsible for a contiguous range $r_a$ of values. Ranges are assigned to different overlay nodes during the initial join process. Under ideal condition, the system guarantees range-based lookups within each routing hub in $\mathcal{O}\log^2 n/k$ when each node maintains $k$ fixed links to the other nodes.

Note that, while the notion of a circular overlay is similar to DHTs, Mercury do not use any randomizing cryptographic hash functions for placing the nodes and data on the overlay. In contrast, Mercury overlay network is organized based on set of links. These links include the: i) successor and predecessor links within the local attribute hub; ii) $k$ links to other nodes in the local attribute hub (intra-hub links) ; iii) one link per hub (inter-hub link) that aids in communicating with other attribute hubs and resolving multi-attribute range queries. Note that, $k$ intra-hubs links is a configurable parameter and could be different for different nodes in the attribute overlay. In this case, the total routing table size at a node is $k + 2$. When a node $n_k$ is presented with message to find a node that mantains a range value $[l_i, r_i]$, it chooses the neighbor $n_i$ such that the clockwise distance $d(l_i, v)$ is minimized, in this case the node $n_i$ maintains the attribute range value $[l_i, r_i]$. Key to message routing performance of Mercury is the choice of $k$ intra-hub links. To set up each link $i$, a node draws a number $x \in \mathcal{I}$ using the harmonic probability distribution function: $p_n(x) = \frac{1}{n \log x}$. Following this, a node $n_i$ attempts to add the node $n^`$ in its routing table which manages the attribute range value $r + (M_a - m_a) \times x$; where $m_a$ and $M_a$ are the minimum and maximum values for attribute $a$. For routing a data record $D$, the system route to the value $\pi_a(D)$. For query $\mathcal{Q}$, $\pi_a(\mathcal{Q})$ is a range. In this case, first the message is routed to the first node that holds the starting range values and then range contiguity property is used to spread the query along the overlay network.

### 5.4.3 PHT: Prefix Hash Tree

The work [89] presents a mechanism for implementing range queries over DHT based system via a trie-based scheme. The bucket in the trie is stored at the DHT node obtained by hashing its corresponding prefixes. The resulting data structure is referred as a trie[1]. In the PHT, every vertex corresponds to a distinct prefix of the data domain being indexed. The prefixes of the nodes in the PHT form a universal prefix set [2]. The scheme associates a prefix label with each vertex of the tree. Given a vertex with label $l$, its left and right child vertices's are labeled as $l_0$ and $l_1$ respectively. The root of the tree is always labeled with the attribute name and all the subsequent vertexes are labeled recursively.

A data item is mapped and stored at the node having longest prefix match with the node label. A node can store upto $B$ items, in case this threshold is exceeded, a node is recursively divided into two child nodes. Hence, this suggests that data items are only stored in the leaf nodes in the PHT and the PHT itself grows dynamically based on distribution of inserted values. This logical PHT is distributed across nodes in the DHT-based network. Using the DHT look-up operation, a PHT node with label $l$ is thus assigned to a node with identifier closest to HASH($l$). Look-up for a range query in PHT network is performed by locating the node corresponding to the longest common prefix in the range. When such a node is found, then parallel traversal of its sub-tree is done to retrieve all the desired items. Note that, significant query look-up speed-up can be achieved by dividing the range into a number of sub-ranges.

### 5.4.4 JXTA: JXTA Search

JXTA Search [114] is an open framework based on the JXTA [58] routing substrate. JXTA search network consists of search hubs, information providers and information consumers. The network message communication protocol is based on the XML format. In the JXTA network, search hubs are organized into $N$ distinct groups. These groups are referred to as *advertisement groups*. These search hubs act as point of contact for providers and consumers. Further each search hub is a member of a network of hubs which has at least one representative of hubs from every advertisement group. These groups are termed as *query groups*. Hence, in this case there is 100% reachability to all stored information in the network.

Every information provider in the network registers its resource information with its local search hub. Each hub periodically sends update message (new additions and deletions of registrations) to all the hub in its advertisement group. In case, the grouping of hubs is content-based, the advertisement is forwarded to the relevant representative for that content. When ever a information consumer wishes to look for a data on the search network, it issues a information request query to the hub it knows or has membership. The hub that receives this query first searches its local index and then other hubs in its advertisement group. If a match is found in the same advertisement group, then the query is forwarded to that hub. In case the query cant be resolved in the local advertisement group then it is broadcasted to all remaining advertisement group using a query group membership information. However, if the search network is organized based on content, then the query is routed to the advertisement group responsible for indexing the desired content.

### 5.4.5 P2PR-Tree: An R-Tree Based Spatial Index for P2P Environments

The work [81] presents a scheme for adopting the R-tree [78] in a P2P setting. P2PR-tree statically divides the multi-dimensional attribute space (universe) into a set of blocks (rectangular tile). The bolcks formed as a result of initial division of space forms the level 0 of the distributed tree. Further, each block is again statically divided into a set of groups, these groups constitute level 1 in the tree. Any further division on the group level (subsequently on the subgroup) is done dynamically and are designated as subgroups at level $i$ ($i \geq 2$). When a new peer joins the system, it contacts one of the existing peer which informs it about the Minimum Bounding Rectangle (MBR) of the blocks. Using this overall block structure information, a peer decides which block(s) belongs to it.

When relevant block(s) is determined, a peer queries any peer in the same block for compiling group-related MBR information. It also queries atleast one peer in every group. Using this group structure information, a peer knows about its group. After determining the group, the same process it utilized for determining the subgroups and so on. Effectively, a peer maintains following routing information: (i) pointers to all blocks in the universe; (ii) pointers to all groups in its block ; (iii) pointer to all subgroups in its group and ;(iv) finally pointers to all peers in its subgroup.

---

[1] A trie is a multi-way retrieval tree used for storing strings over an alphabet in which there is one node for every common prefix and all nodes that share a common prefix hang off the node corresponding to the common prefix.

[2] A set of prefix is a universal prefix set if and only if for any infinite binary sequence $b$ there is exactly one element in the set which is a prefix of $b$

The scheme defines threshold value on maximum number of peers in a group and subgroup denoted by $G_{Max}$ and $SG_{Max}$.

A query $Q_L$ for a object is propagated recursively top down starting from the level 0. As the query arrives at any peer $P_i$ in the system, $P_i$ checks whether its MBR covers the region indexed by the query. If yes, then $P_i$ searches its own R-tree and returns the results and the search is terminated at that point. Else the peer forwards the query to relevant block or group or subgroup or peer using its routing table pointers. This process is repeated untill query block is located or the query reaches dead end of the tree.

# 6 Open Issues

Current models of distributed systems including the Grid computing and P2P computing suffers from a knowledge and resource fragmentation problem. By knowledge fragmentation, we mean that various research groups in both academia and industry work in a independent manner. They define standards without any proper coordination. They give very little attention to the inter-operatibility between the related systems. Such disparity can be seen in the operation of various grid systems including Condor-G, Nimrod-G, OurGrid, PlanetLab, Grid-Federation, SETI@Home, Tycoon and Bellagio. These systems define independent interfaces, communication protocols, superscheduling and resource allocation methodologies . In this case users have access to only those resources that can understand the underlying Grid system protocol. Hence, leading to distributed resource fragmentation problem. These systems do not expose any API or interfaces that can help them to inter-operate. In other words, a user from Condor-G grid can not submit his job to a Tycoon grid etc. A possible solution to this can be federating these grid systems based on universally agreed standards (similar to TCP/IP model that governs the current Internet). The core to the operation and interoperability of Internet component is the common resource indexing system i.e. DNS. Both Grid and P2P community clearly lacks any such global or widely accepted service.

Possible bottleneck to realizing such large scale federation includes: (i) availability of a robust, distributed, scalable resource indexing/organization system; (ii) evolution of common standards for resource allocation and application superscheduling; (iii) agreement on using common middle-wares for managing grid resources such as clusters, SMPs etc; and (iv) defining common interfaces and APIs that can help different related system to inter-operate and coordinate activities. However, focus of this article is on designing a robust, distributed and scalable resource indexing system that can enable such large scale federated environment.

To this end, we outlined various approaches currently being proposed for solving Grid resource indexing problem. Every approach has its own merits and limitations. Some of these issues we highlighted in the resource and P2P network organization taxonomy section. However, few questions still remain to be answered including: (i) whether current approaches provide ample flexibility in query composition and resolution that is required to undertake complexity of Grid superscheduling systems; (ii) is there a need to define a new query composition language which is capable of representing all possible kinds of queries ambient now or that can arise in future; and (iii) whether the current DHT based P2P techniques are robust enough to support such complex query systems.

# 7 Summary and Conclusion

Recent past has observed increase in complexities involved with grid resources including their management policies, organization and scale. Key elements that differentiate a computational grid system from a PDCS includes: (i) autonomy; (ii) decentralized ownership; (iii) heterogeneity in management policies, resource types and network interconnect; and (iv) dynamicity in resource conditions and availability. Traditional grid systems [53, 18, 12] based on centralized information services are proving to be bottleneck as regard to scalability, fault-tolerance and mechanism design issues. To address this, P2P based resource organization is being advocated. P2P organization is scalable, adaptable to dynamic network conditions and highly available.

In this work, we presented a detailed taxonomy that characterizes issues involved in designing a P2P/decentralized GRIS. We classified the taxonomies into two sections: (i) resource taxonomy; and (ii) P2P taxonomy. Our resource taxonomy highlighted the attributes related to a computational grid resource. Further, we summarized different kinds of queries that are ambient in current computational grid systems. In general, Grid superscheduling query falls under the category of multi-dimensional point or window query. However, it still remains to be seen whether a universal grid resource query composition language is required to express different kinds of Grid RLQs and RUQs.

We presented classification of P2P approaches based on three dimensions including: (i) P2P network organization; (ii) approaches to distribution of the data among the peers; and (iii) routing of queries. In principal, data distribution mechanism directly dictates how the query is routed among the relevant peers. Multi-dimensional resource indexing data is distributed among peers by utilizing the data structures such as SFCs, quad-trees, R-tree and KD-tree. Some approaches also modified existing hashing scheme to facilitate the single-dimensional range queries in DHT network.

However, it still remains to be analyzed which of the surveyed approach is best suited for organizing a GRIS. We also surveyed some of the existing grid resource indexing techniques and classified them into different categories using the proposed taxonomy.

# 8   ACKNOWLEDGMENTS

Firstly, we shall like to acknowledge the authors of the papers whose work have been surveyed and utilized in developing taxonomy in this paper. We thank our group members at the University of Melbourne- Marcos Assuncao, Al-Mukaddim Khan Pathan, Hussein Gibbins, Krishna Nadiminti, Md Mustafizur Rahman - for their comments on this paper. We shall also like to thank Adriana Iamnitchi (University of South Florida) for her valuable feedbacks on the initial version of this work.